\documentclass{aa}  

\usepackage{graphicx}
\usepackage{txfonts}
\usepackage{lipsum}
\usepackage{subcaption}       
\usepackage{hyperref}

\usepackage{xcolor}
\definecolor{Orange}{rgb}{1.0,0.5,0.15}
\definecolor{Blue}{rgb}{0,0.08,0.65}
\definecolor{Red}{rgb}{0.65,0.08,0.05}
\definecolor{Green}{rgb}{0.15,0.45,0.25}
\definecolor{Pink}{rgb}{1.0,0.05,0.5}
\definecolor{bubbles}{rgb}{0.91, 1.0, 1.0}
\definecolor{aquamarine}{rgb}{0.5, 1.0, 0.83}
\definecolor{bubblegum}{rgb}{0.99, 0.76, 0.8}
\definecolor{bluebell}{rgb}{0.74, 0.74, 0.92}
\definecolor{dollarbill}{rgb}{0.72, 0.93, 0.6}

\begin{document}
   \title{Dissecting the Perseus-Pisces supercluster observed with CFHT-MegaCam}

    \subtitle{Investigating environmental effects on galaxy morphology}
    
\author{ M.~Mondelin
\inst{1}
 \and
    S.~Codis
    \inst{1}
    \and
    J.-C.~Cuillandre
    \inst{1}
    \and
    C. Laigle
    \inst{2}
    \and
    A. Boselli 
    \inst{3, 4}
   \and
  K. Kraljic
  \inst{5}
  \and
    C. Stone
    \inst{6}
   }

   \institute{ Universit\'e Paris-Saclay, Universit\'e Paris Cit\'e, CEA, CNRS, AIM, 91191 Gif-sur-Yvette, France \label{1}\\ \email{maelie.mondelin@cea.fr} 
   \and
    Institut d’Astrophysique de Paris, UMR 7095, CNRS, and Sorbonne Université, 98 bis boulevard Arago, 75014 Paris, France \label{2}
   \and
    Aix-Marseille Universit\'e, CNRS, CNES, LAM, Marseille, France\label{3}
    \and
    INAF - Osservatorio Astronomico di Cagliari, Via della Scienza 5, 09047 Selargius (CA), Italy\label{4}
    \and
    Observatoire Astronomique de Strasbourg, Université de Strasbourg, CNRS, UMR 7550, 67000 Strasbourg, France\label{5}
    \and
    Department of Physics, Universit\'{e} de Montr\'{e}al, 2900 Edouard Montpetit Blvd, Montr\'{e}al, Qu\'{e}bec H3T 1J4, Canada\label{6}
    }

    \date{}
\abstract{
{The discovery of the large-scale structure of the Universe has fundamentally reshaped our understanding of galaxy formation and evolution. Filaments of the cosmic web act as privileged environments, guiding the growth of structures.}

{Motivated by predictions from cosmological simulations, we investigate the morphological distribution of galaxies within the Perseus-Pisces Supercluster, a prominent nearby filamentary structure at $\sim$70 Mpc. In particular, we examine how galaxy morphology and structural perturbations correlate with location in the filament network and with proximity to dense nodes.}

{We construct a galaxy sample from a spectroscopic catalogue cross-matched with deep $r$-band CFHT/MegaCam imaging from the Ultraviolet Near-Infrared Optical Northern Survey and additional observations, enabling the exploration of low surface brightness features and the detection of extended outer structures. Galaxy morphologies are classified both visually and via structural parameters derived from surface brightness profiles, using the \texttt{AutoProf} and \texttt{AstroPhot} Python tools. The 3D filamentary skeleton of the Perseus-Pisces Supercluster is extracted using the \texttt{DisPerSE} algorithm, allowing the computation of the perpendicular distance of each galaxy to the nearest filament ($d_{\mathrm{fil}}$) and to group/cluster centers ($d_{\mathrm{gr,cl}}$).}

{The 3D reconstruction uncovers a network of interconnected sub-filaments converging around the Pisces cluster, forming a complex, multi-branched supercluster-scale structure that likely governs environmental influences on galaxy evolution. We find clear evidence of morphological and stellar mass segregation within the Perseus-Pisces Supercluster: massive early-type galaxies (E/S0) predominantly reside along the spine of filaments and in the vicinity of dense nodes, whereas late-type (S/Irr) systems are more diffusely distributed throughout the network. Approximately $10-13\%$ of galaxies display pronounced signatures of gravitational interaction, with stellar halo asymmetries being most common in filaments and groups. Together, these results highlight the dual role of filamentary environments, both as sites hosting evolved early-type systems and as regions where local tidal interactions and pre-processing significantly affect galaxy morphology.}
}


\titlerunning{The Pisces-Perseus supercluster observed with CFHT-MegaCam}
\authorrunning{Mondelin et al.}

   \maketitle
 
\section{Introduction}

The large-scale structures of the Universe, commonly referred to as the cosmic web, provide the fundamental framework for understanding the formation and evolution of galaxies across cosmic time. In the standard $\Lambda$CDM cosmological model, the present-day matter distribution results from the gravitational growth of tiny initial density fluctuations imprinted in the early Universe. This process leads to the emergence of a complex network of nodes, filaments, sheets, and voids, collectively shaping the cosmic web \citep{Klypin1983,Bond1996}.
The anisotropic collapse of matter, well described by Zel’dovich’s perturbation theory \citep{Zeldovich1970}, governs the formation of these structures, which are now extensively studied through large cosmological simulations incorporating both dark matter dynamics and baryonic physics \citep[e.g.,][]{Eagle1_2015, Kaviraj2017, Illustris}.

Observationally, large galaxy redshift surveys such as the CfA Redshift Survey \citep{Huchra1983} and the Sloan Digital Sky Survey (SDSS, \cite{SDSS2000}) have revealed and validated the filamentary nature of the galaxy distribution, providing stringent constraints on cosmological models. These surveys have demonstrated that galaxy properties such as morphology, star formation activity and stellar mass are strongly influenced by their location within the cosmic web. In particular, a well-established morphological segregation is observed: early-type galaxies preferentially reside in dense nodes and cluster cores, whereas late-type, star-forming spiral galaxies dominate lower-density filamentary and field environments \citep{Dressler1980,Dressler1997,Tempel2011,Buta2017,2020MNRAS.491.4294K}.

\begin{figure}
\centering
\includegraphics[width=\columnwidth]{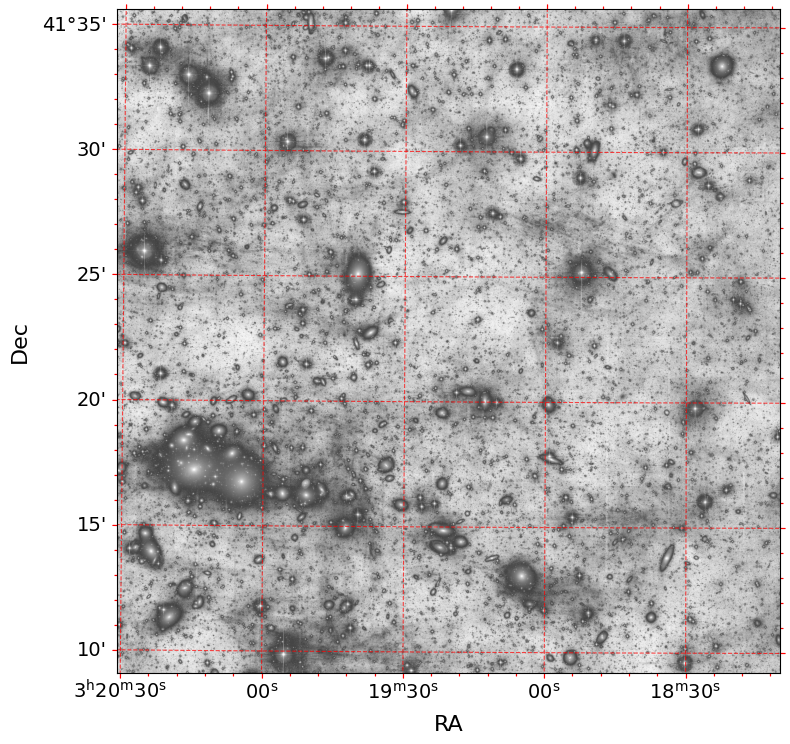}
\caption{Observation tile in the CFHT/Megacam $r$-band band covering a region of 1.2 deg$^2$ near the Perseus cluster (visible in the lower left corner of the image), optimized for low surface brightness detection (3×3 binning).}

\label{fig:tileobs}
\end{figure}
As they evolve, galaxies embedded in massive dark matter halos form in small overdensities and migrate along filaments, gradually assembling into groups and clusters. Throughout this journey, they interact with their surroundings \citep[e.g][]{ATTT,2018MNRAS.476.4877M}, undergoing processes that can significantly alter their morphology and star formation activity. 
This environmental dependence arises from a complex interplay between different physical mechanisms such as tidal interactions \citep{Toomre1972, Duc2013}, ram-pressure stripping \citep{Gunn1972, Boselli2006,Poggianti2017} or galaxy mergers \citep{Toomre1972, Lotz2008}, which collectively may quench star formation and drive morphological transformation, especially at low redshift, where environmental effects become increasingly significant \citep{Boselli2006, Ilbert2013}.

To fully understand how galaxies evolve within the cosmic web, particularly under the influence of their large-scale environment, high-resolution and deep imaging surveys are essential. While significant progress has been made thanks to low-redshift observations from surveys like the SDSS \citep{Jones2010,Martinez2016}, theoretical insights from hydrodynamical simulations such as Horizon-AGN \citep{Dubois2014} and Illustris-TNG \citep{Nelson2018} underscore the importance of capturing faint, extended structures and subtle morphological features that trace environmental interactions.

Looking ahead, the next generation of wide-field cosmological surveys, particularly the \emph{Euclid} survey and the Legacy Survey of Space and Time (LSST), are poised to revolutionize our view of galaxy evolution by mapping the cosmic web across vast volumes and over a wide range of redshifts \citep{EuclidSkyOverview,LSST2009}. Early science results from \emph{Euclid}’s first data release \citep{Aussel2025} already highlight the value of deep, multi-band imaging in probing galaxy alignments and morphological transformations linked to environment. For instance, studies in the \emph{Euclid} Deep Fields have revealed correlations between galaxy shapes and the surrounding large-scale structure \citep{Laigle2025}, along with strong environmental imprints in cluster galaxies \citep{Gouin2025}.

In this context, nearby large-scale structures such as the Perseus-Pisces filament serve as ideal laboratories for studying environmental processes in detail and anchoring interpretations of high-redshift observations. The Perseus Pisces filament is a particularly compelling example, as it is a 
coherent and elongated structure that connects multiple clusters and groups and was initially identified through HI 21 cm surveys and optical redshift catalogs \citep{Gregory1978,Chincarini1983,Giovanelli1986,Ramatsoku2017,Bohringer2021}. Located at a distance of approximately 70 Mpc ($z=0.017$), spanning nearly 60 degrees on the sky (comoving volume of roughly $3.4\times10^{5}\,\mathrm{Mpc}^{3}$), the Perseus Pisces supercluster (hereafter PPSC) was one of the first superclusters to be discovered \citep{Gregory.1981}. 
Previous studies have documented morphological segregation within this filament, with early-type galaxies populating denser regions \citep{Chincarini1983,Giovanelli1986}. Observations also provided evidence of environmental processes such as ram pressure stripping occurring in embedded cluster cores \citep{Skillman1996,Moriondo1998,Moriondo2001,vanGorkom2004, Roberts2022}, and recent \emph{Euclid} Early Release Observations (ERO) of the Perseus region \citep{ George2025} revealed a coexistence of mechanisms that collectively shape galaxy evolution in these dense and dynamic environments \citep{EROPerseusOverview}.

Building on this, we exploit a unique, deep imaging dataset from the Canada-France-Hawaii Telescope (CFHT), partially obtained as part of the Ultraviolet Near Infrared Optical Northern Survey (UNIONS; \citealt{Savary2022}), and acquired with the MegaCam instrument. 
Thanks to its exceptional sensitivity to low surface brightness, reaching down to approximately $\sim$29 mag arcsec$^{-2}$ in the $r$-band, MegaCam is capable of detecting extremely faint structures. 
This includes tidal features and extended stellar halos that are predominantly composed of old, red stellar populations. We propose to investigate the influence of the large-scale environment on galaxies properties within this particular large-scale structure, the PPSC, as observed by MegaCam. An example of such an observation is presented in Fig.~\ref{fig:tileobs}, which shows a representative a $r$-band tile near the Perseus cluster, optimized for low surface brightness detection.

This paper is organized as follows. In Sect.~\ref{sc:Method}, we describe the methodology used to construct our galaxy catalogue within the regions of interest, extract the filamentary structure of the PPSC, and derive photometric and physical properties for galaxies from the dataset. In Sect.~\ref{sc:Results}, we present the diversity of interaction features observed across the filament and characterize the morphological segregation within the structure. Finally, Section~\ref{sc:Discussion} discusses the implications of our findings for the evolutionary history of galaxies in this prominent nearby supercluster.

\section{Method}\label{sc:Method}

To characterize the large-scale environment around the PPSC, we first describe the dataset used to trace the surrounding cosmic web. A wide-field, spatially extended galaxy catalogue is constructed, covering a broad sky area and redshift range that fully encompass both
observational regions. This extensive coverage enables an unbiased reconstruction of the local cosmic web and mitigates edge effects that could otherwise distort the topology of large-scale structures. From this, we extract a filamentary skeleton that provides a physically motivated and statistically robust framework for our environmental analysis. The resulting catalogue serves also as the fundamental dataset to derive a physically meaningful and statistically robust filamentary skeleton that underpins our subsequent environmental analyses.

\subsection{Observational data}

\begin{figure*}
\centering
\includegraphics[width=\textwidth]{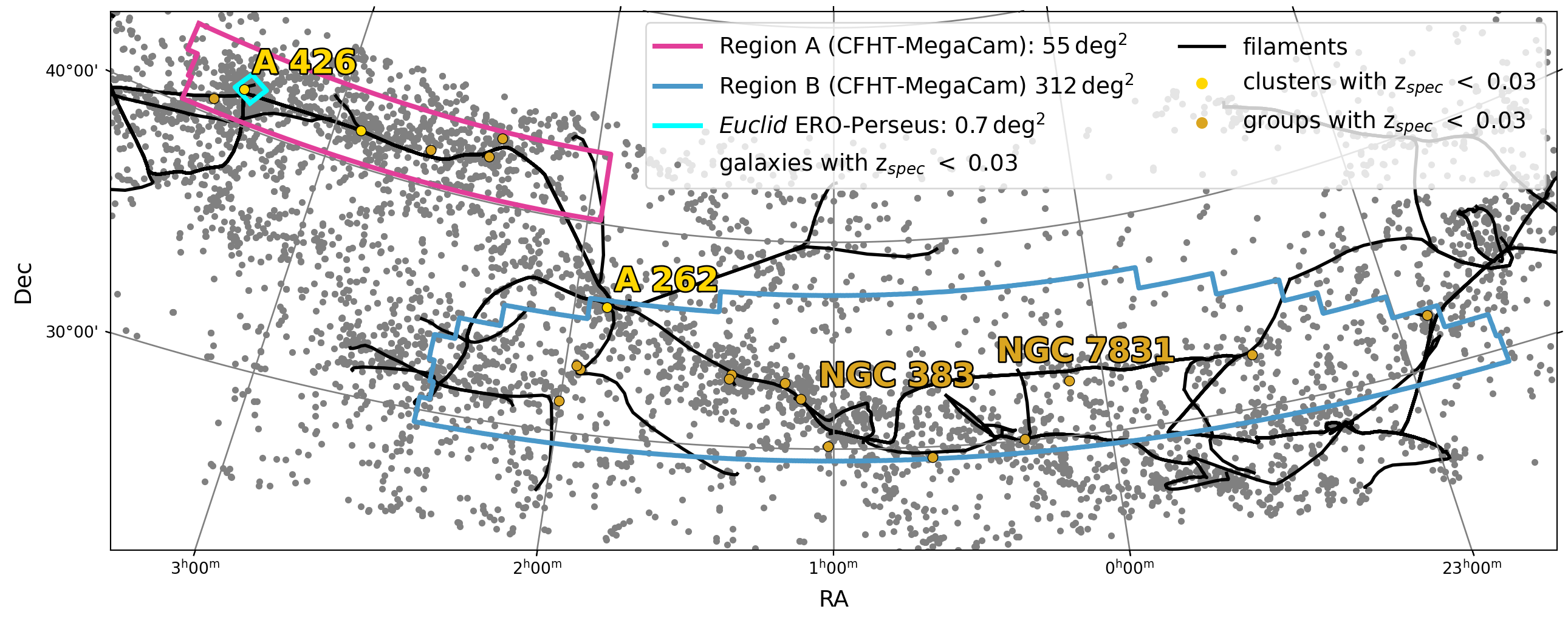}
\caption{Sky-projected distribution of galaxies in the PPSC catalog. Grey dots represent galaxies from the PPSC catalog. Group and cluster centres identified are shown as gold points: from the left to the right in the region A, A\,426, AWM7, a group with UGC\,1841, and A\,347. The projected skeleton, extracted from our final 3D structure, is represented by dark lines and traces the projected filamentary network. The boundaries of the region A and the region B are shown in purple and blue, respectively. A cyan contour indicates the footprint of the Euclid ERO Perseus field, overlaid for reference.} 
\label{2D_skeleton}
\end{figure*}

The footprint of our analysis consists of two contiguous subregions of the PPSC, referred to as regions A and B. The imaging data were obtained with the MegaCam instrument on the Canada–France–Hawaii Telescope. Region B, shown in blue in Fig.~\ref{2D_skeleton}, covers an area of 310~deg$^{2}$ observed as part of the Ultraviolet Near-Infrared Optical Northern Survey in the South Galactic Cap\footnote{\url{https://unions.skysurvey.cc/Publications/Policies}}. It extends from the north-western part of the field toward the Pisces cluster at lower declinations. Region A, outlined in pink in Fig.~\ref{2D_skeleton}, spans a complementary 52~deg$^{2}$ observed through an independent but comparable programme. It is centred on the prominent galaxy clusters A\,426 (Perseus), AWM7, and A\,347, as well as a nearby group hosting the central galaxy UGC\,1841. Both datasets were acquired in the $r$-band and reach similar depths, allowing a consistent joint analysis across the full structure. The total coverage of 362~deg$^{2}$ was obtained under dark skies with sub-arcsecond seeing (median 0.7$\arcsec$), reaching a point-source depth of $r = 25.1$ at 5$\sigma$. MegaCam provides a 1.1~deg$^{2}$ field of view with a native resolution of 0.187\arcsec~per pixel. Each pointing was observed with three 2-minute exposures and processed using the \texttt{Elixir-LSB} pipeline \citep{Ferrarese2012}, optimised for the detection of low surface brightness features (Fig.~\ref{fig:tileobs}).

The two areas are clearly delineated in the two-dimensional projection presented in Fig.~\ref{2D_skeleton}, which serves as a reference for the spatial extent and location of the regions studied.

\subsection{The PPSC catalogue and the skeleton of the PPSC}

We assemble a spectroscopic PPSC catalogue by merging data from multiple extragalactic surveys and databases, including HyperLEDA\footnote{\url{http://leda.univ-lyon1.fr/}} \citep{Makarov2014}, NED\footnote{\url{https://ned.ipac.caltech.edu/} The NASA/IPAC Extragalactic Database (NED) is operated by the Jet Propulsion Laboratory, California Institute of Technology, under contract with the National Aeronautics and Space Administration.}, the 2MASS Redshift Survey (2MRS; \citealt{Huchra2012}), the Sloan Digital Sky Survey (SDSS;\citealt{SDSS2000}) and the FAST all sky \ion{H}{i} survey (FASHI; \citealt{Zhang_2023}).Although the sample is not strictly magnitude-complete, it provides the most extensive spectroscopic coverage currently attainable in these environments.

To isolate the PPSC structure, we select galaxies within the sky coordinates RA $\in [22^h, 4^h]$ and Dec $\in [+25^\circ, +45^\circ]$, and spectroscopic redshift range $z \in [0.004, 0.03]$, corresponding to approximate distances between 17 and 127 Mpc. We include only galaxies with spectroscopic redshifts to ensure an accurate 3D representation. This yields a coherent three-dimensional structure containing 7\,351 galaxies that defines the foundation for subsequent structural, photometric, and morphometric analysis. While extensive, this compilation is not complete in stellar mass, as it relies exclusively on galaxies with available spectroscopic redshifts. The implications of this incompleteness are discussed in Sect.~\ref{subsc:mass}. The celestial coordinates and redshifts are converted into comoving Cartesian coordinates assuming a flat $\Lambda$CDM cosmology consistent with \citet{Planck2018} and \citet{Planck2020}.

To recover the intrinsic three-dimensional structure of the PPSC, we first correct for redshift-space distortions, particularly the Fingers-of-God effect (FoG hereafter), using a Friends-of-Friends (FoF) algorithm \citep{Kraljic2018}. Galaxies are linked using a redshift difference of $\Delta z \leq 0.005$ and an angular separation of $< 0.4^\circ$, with a minimum group richness of ten members. These parameters were empirically determined to provide a balance between effectively collapsing virialised structures and avoiding the artificial merging of distinct systems. This choice was validated through visual inspection of the resulting group distributions.  

Residual distortions after correction are minimal across most of the survey area and primarily affect the Perseus cluster, where the high velocity dispersion leads to elongated structures along the line of sight. These residuals, however, have a negligible impact on the overall filament geometry. From this point onward, galaxy positions refer to their coordinates after correction for FoG effects.

We then construct the filamentary skeleton using a robust statistical approach based on multiple realizations, inspired from the methodology introduced in the 2D analysis of \citet{Laigle2025}. Specifically, we generate 100 bootstrap realizations of the galaxy distribution, where in each realization 5\% of galaxies are randomly removed to account for sampling variance. For each of these realizations, we apply the Discrete Persistent Structure Extractor code (\texttt{DisPerSE}; \citealt{Sousbie2011, Sousbie2011b, sousbie2011direct}), which identifies topological structures in the discrete galaxy field based on persistent homology. The skeleton is extracted with a persistence threshold of $3\sigma$ to filter out spurious features and includes two topological smoothing iterations of the skeleton. These parameters were chosen empirically to best reveal the most massive and coherent structures within the PPSC volume.

From these realizations, we build a 3D filament traversal density grid, which quantifies how frequently each voxel is crossed by a filament. This grid is used as input for a final \texttt{DisPerSE} run (with the same parameters), yielding a robust skeleton that we will use for the rest of our analysis. A detailed description of the skeleton construction methodology, including intermediate steps and validation figures, is provided in Appendix~\ref{app:skeleton}. 

\begin{figure}
\centering
\includegraphics[width=\columnwidth]{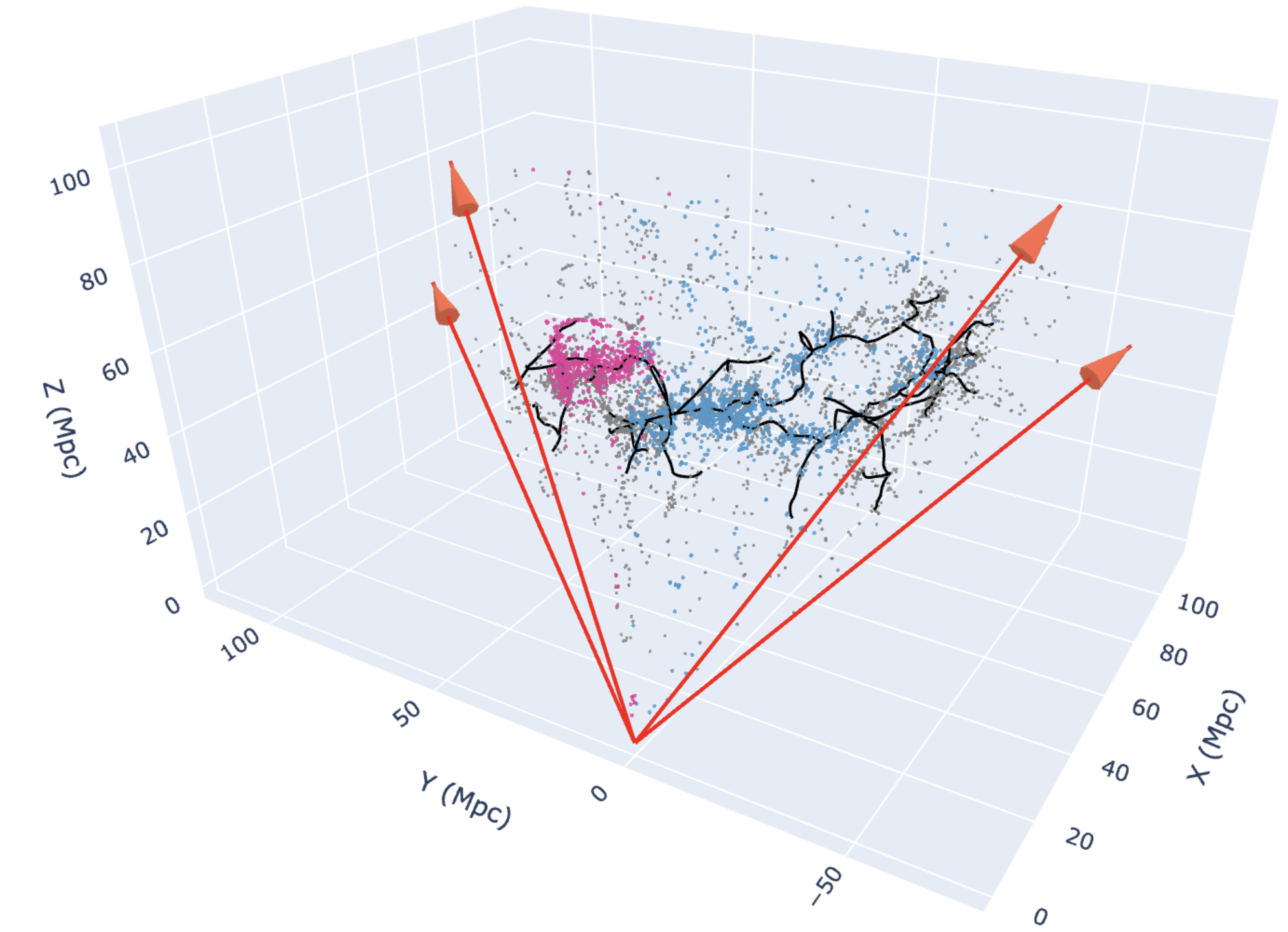}
\caption{Three-dimensional reconstruction of the PPSC. Grey dots represent galaxies from the PPSC catalog, pink dots correspond to galaxies in the region A, and blue dots indicate galaxies in the region B. The final filamentary skeleton, shown as dark lines, is extracted using \texttt{DisPerSE} from a filament density grid built from multiple realizations of the galaxy distribution. The red arrows mark the cone of interest and highlight the directions of increasing redshift.}
\label{3D_skeleton}
\end{figure}

The resulting 3D skeleton consists of 73 filaments and connected through 70 nodes (Figure~\ref{3D_skeleton}). 

The sky-projected view of the reconstructed structure is shown in Fig.~\ref{3D_skeleton}, where the dark lines correspond to the projection of the three-dimensional filamentary skeleton onto the celestial sphere. For illustration purposes, we also derived a two-dimensional skeleton directly from the projected galaxy distribution, in order to evaluate the impact of deprojection effects.

The two regions of interest previously described have the following organization: 
\begin{itemize}
    \item The region A, outlined in pink in Fig.~\ref{2D_skeleton} and including 915 galaxies, contains a high-density node encompassing the Perseus cluster along with two other significant galaxy concentrations. This region is highly zoomed-in around these clusters, offering a detailed view of these dense environments and their immediate surroundings.
    \item The region B, shown in blue in Fig.~\ref{2D_skeleton}, corresponds to a coherent, linear structure tracing a prominent elongated filament of the cosmic web. This filamentary structure is also illustrated in three dimensions in Fig.~\ref{3D_skeleton}, providing a volumetric perspective of its spatial complexity. It extends from the northwestern part of the surveyed area toward the Pisces cluster at lower declination. An apparent extension branches off toward the western edge of the field, while another arm emerges near the Pisces cluster and extends toward decreasing right ascension. This dataset contains 1761 galaxies.
\end{itemize}

For each galaxy, we compute the perpendicular distance to the nearest filament segment, denoted as $d_{\mathrm{fil}}$, following the definition of \citet{Laigle2025}. This distance metric serves as a quantitative tracer of a galaxy’s position within the cosmic web. To estimate the uncertainty on $d_{\mathrm{fil}}$, we measure the dispersion of the nearest-distance values obtained across the 100 Monte Carlo realizations of the skeleton, which provides a statistical estimate of positional uncertainty due to sampling.

\subsection{Identification of group and cluster centres}

Groups and clusters were identified within the two CFHT-observed regions. Candidate systems were first extracted from a crossmatch between the group catalogue of \cite{Lu2016}, the clusters and groups catalogue of \cite{Bohringer2021} and NED database. Each candidate system was visually inspected using CFHT imaging to verify the presence of a genuine galaxy overdensity, using a cutout of $1000 \times 1000$ pixels (corresponding to approximately $0.2 \times 0.2$ Mpc$^2$ at the cluster redshift). Only rich systems with at least five confirmed member galaxies were retained, as our analysis focuses on the principal groups and clusters.

The characteristic radius, $r_{200}$, of each system was estimated either directly from the literature catalogue when available, or computed from the redshift of the central galaxy, corrected for the effects of Fingers-of-God, and the line-of-sight velocity dispersion. This provides a physically motivated estimate of the radius within which the mean enclosed density is 200 times the critical density. Uncertainties on $r_{200}$ arise primarily from the statistical errors on the velocity dispersion and, when applicable, from the uncertainties on the corrected redshift of the central galaxy. These propagate into the derived cluster-centric distance $d_{\mathrm{cl,gr}}$.

\subsection{Surface brightness analysis}

To study the outer morphological structures and low surface brightness features of galaxies across environments, we extract square cutouts centered on each galaxy. Standard cutouts of $9.4\times9.4~\mathrm{arcmin}$ ($\approx200\times200~\mathrm{kpc}^2$ at the mean distance of the PPSC) are used to enhance the signal-to-noise ratio in the outskirts, while very extended or overlapping galaxies are assigned larger cutouts of $18.7\times18.7~\mathrm{arcmin}$ ($\approx400\times400~\mathrm{kpc}^2$).
 Galaxies affected by image artefacts, primarily those with a bright star along the line of sight (common due to the low Galactic latitude of our survey), or located near field edges are removed for the further analysis: 30 galaxies are excluded in the region B, and 12 in the region A.

We adopt two complementary surface brightness modelling approaches, depending on the degree of isolation of each galaxy. It is worth noting that this methodology is closely aligned with the one employed in \cite{EROPerseusOverview} and \cite{Mondelin2025}. 

\subsubsection{AutoProf: isolated galaxies}

\begin{figure*}
\centering
\includegraphics[width=\textwidth]{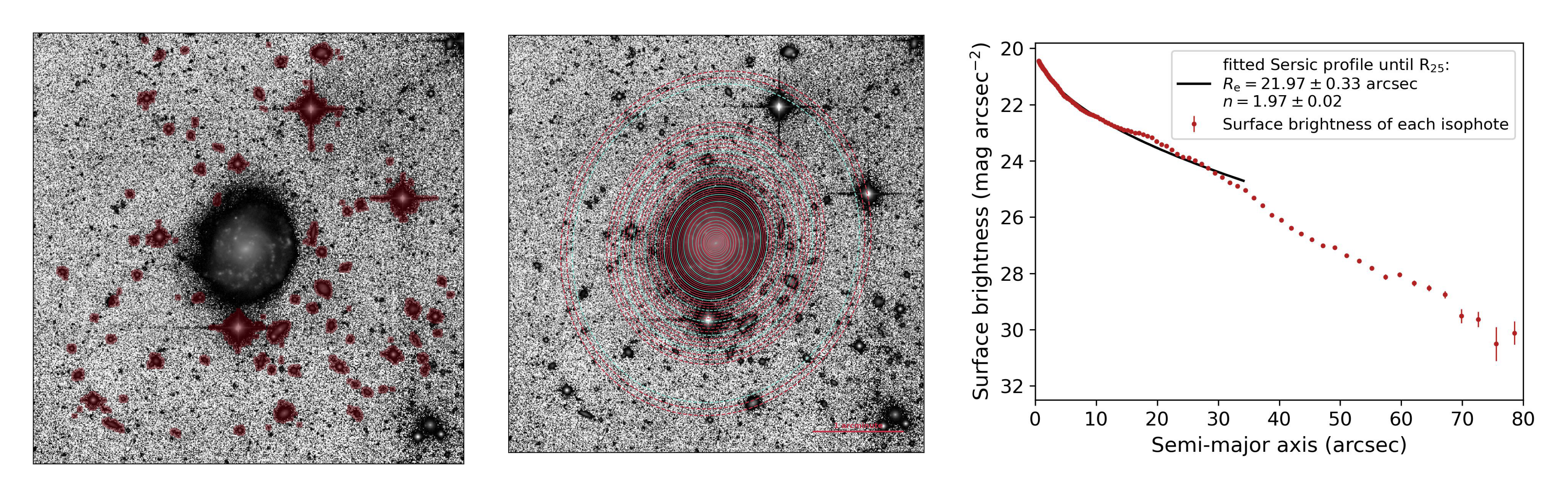}
\caption{Example of \texttt{AutoProf} processing for an isolated galaxy, UGC\,0742, in the region B. Left: masked image with segmentation overlaid. Middle: 2D isophotal extraction. Right: radial surface brightness profile with Sérsic fit out to $R_{25}$. The profiles reach $\sim$29--30~mag~arcsec$^{-2}$, which is representative of the depth achieved for the majority of galaxies in our dataset.}
\label{AutoProf_example}
\end{figure*}

Galaxies classified as isolated, i.e. with no significant overlapping neighbors within their extended envelopes, are analyzed using \texttt{AutoProf} \citep{Stone2021}, a Python-based tool designed to derive azimuthally averaged surface brightness profiles from galaxy imaging data. This method is particularly well-suited for systems with regular, symmetric morphologies where environmental contamination is minimal.

For each galaxy, we generate a segmentation map from the cutouts using \texttt{SExtractor} \citep{Bertin1996}, which identifies all sources above a 1.5$\sigma$ threshold and larger than 5 pixels in area. These sources, which are typically stars, background galaxies, or small companions, are then masked prior to profile extraction.

\texttt{AutoProf} computes elliptical isophotes centered on the galaxy and extracts the azimuthally averaged radial surface brightness profile out to the isophotal radius corresponding to a surface brightness of 29 mag arcsec$^{-2}$, commonly referred to as $R_{29}$.  

It also estimates the local background level as part of the fitting process, as described by \citep{Stone2021}. The resulting structural parameters (e.g., concentration, central brightness, radial extent) provide us with a basis for preliminary morphological classification (e.g., disk-like vs. spheroidal systems).

The diagnostic output of \texttt{AutoProf} includes: (i) the masked image with detected sources overlaid, (ii) the map of elliptical isophotes, and (iii) the radial surface brightness profile, which we then fit with a Sérsic model using the \texttt{curve\_fit} Python function, as illustrated in Fig.~\ref{AutoProf_example}.

This procedure was applied to the majority of galaxies in both the region B and region As, excluding systems with substantial envelope overlap or strong residual contamination from neighboring objects.

\subsubsection{AstroPhot: overlapping systems}
For galaxies located in crowded environments, particularly in cluster cores and compact groups, where outer stellar envelopes overlap significantly, the \texttt{AutoProf} approach proves unreliable due to contamination in the radial profiles. In these cases, we use \texttt{AstroPhot} \citep{Stone2023}, a \texttt{python} tool designed for photometric decomposition in complex fields, capable of modelling multiple sources either simultaneously or in localized windows, as used in \cite{EROPerseusOverview}.

We apply the same masking strategy as for isolated galaxies, but on larger cutouts of $18.7\times18.7~\mathrm{arcmin}$ tiles to account for the extended surroundings. Each target galaxy is modelled individually using a single Sérsic component. Note that while a double Sérsic model would be more appropriate for spiral/S0 galaxies, a single component provides sufficiently accurate photometry within $R_{25}$ for our purposes, given that most targets are early-type systems in dense environments. We also initialise a simple constant model to account for the local background level.

The fitting procedure is performed using a maximum likelihood estimator. We visually inspect the residual maps to assess the quality of the fit. This method enables a robust extraction of structural parameters, even in the presence of overlapping light profiles or diffuse background emission, i.e. galactic cirrus. 

\begin{figure*}
\centering
\includegraphics[width=\textwidth]{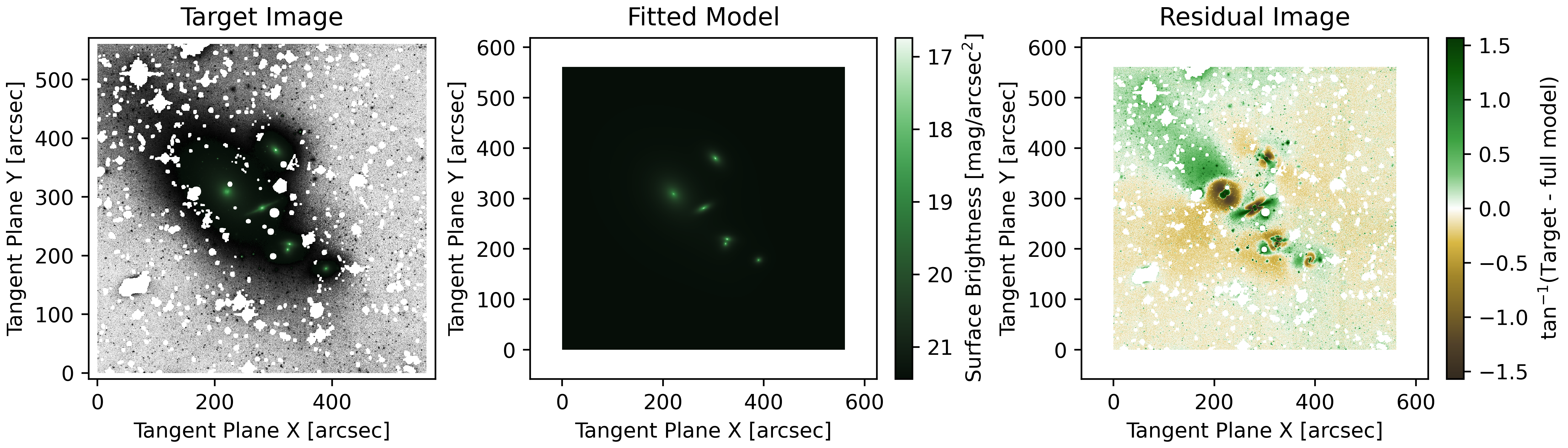}
\caption{Example of \texttt{AstroPhot} modelling in a crowded region. Sérsic profiles are fitted to individual galaxies from the PPSC catalog, displayed on this $2000 \times 2000$ pixel tile.}
\label{AstroPhot_example}
\end{figure*}

This approach is used for 25 galaxies in the region B and 32 in the region A. Figure~\ref{AstroPhot_example} shows a representative example of a galaxy fitted with \texttt{AstroPhot}, including both the model and the residual image.

\subsubsection{Morphology identification}

In total, our catalogue contains 903 galaxies in region A, including 36 compact sources with available spectroscopic redshifts that are not considered in the following analysis focused on massive galaxies. Region B contains 1849 galaxies, of which 118 spectroscopic sources are similarly excluded. The final working samples therefore comprise 867 galaxies in region A and 1731 galaxies in region B.

Alongside the structural analysis, we visually inspect all galaxies using the $r$-band cutouts to identify signs of dynamical interactions. Features are flagged according to visual inspection, as illustrated in Fig.~\ref{interaction_examples}. We adopt a broad classification consistent with simulations and previous observational studies: ram-pressure stripping, recognized through sharp truncations or one-sided extraplanar features in the disk and halo \citep{Gunn1972, Boselli2006,Poggianti2017}; ongoing major mergers, identified by the presence of double nuclei, strong distortions, or bridges connecting companion galaxies \citep{Toomre1972, Lotz2008}; tidal tails, which trace major mergers and appear as broad extensions of the host \citep{Toomre1972, Duc2013}; and shells, mainly produced in radial encounters, appearing as arc-like overdensities \citep{Quinn1984}. In addition, we flag asymmetries, for which we examine the outer halo beyond $R_{27}$ (isophotal radius corresponding to a surface brightness of 27 mag arcsec$^{-2}$) to search for deviations in shape or centroid offsets relative to the galaxy body. This step is performed visually, focusing on large-scale halo asymmetries rather than substructures, and is not applied to galaxies with irregular morphology, since they are by definition asymmetric with no bulge.

Each galaxy is assigned a morphological type based on its visual appearance and the shape of its surface brightness profile, following the standard Hubble sequence \citep{Hubble1926, deVaucouleurs1959, Sandage1961, Huertas-Company2011}. The classification distinguishes between elliptical (E; $T \leq -5$), lenticular (S0; $-5 < T \leq -1$), spiral (S(B)a–S(B)d; $0 \leq T \leq 9$), and irregular (Irr; $T = 10$) galaxies. Visual inspection is complemented by the analysis of the Sérsic index and global structural parameters, which help to distinguish between disc- and bulge-dominated systems \citep{Sersic.1963}. Elliptical systems are identified by their smooth, centrally concentrated light profiles and high Sérsic indices ($n \gtrsim 3.5$). S0 galaxies exhibit a prominent bulge with a faint disc component and intermediate Sérsic indices ($n \sim$2-3). Spiral galaxies, associated with lower Sérsic indices ($n \lesssim 2$) and higher ellipticities, display clear disc structures with spiral arms and a visible bar or a bulge. Irregular galaxies show asymmetric shapes lacking ordered structures.

Representative examples of these morphological types, as identified through visual inspection and Sérsic profile shape, are shown in Fig.~\ref{morpho_examples}. Note that Appendix \ref{app:sclingrelations} presents the scaling relation illustrating the distribution across several parameters.

\subsubsection{Stellar mass estimation and completeness}\label{subsc:mass}

Stellar masses for galaxies in our sample were estimated following the model from \citet{Bell2003} with a Chabrier initial mass function \citep{Chabrier2003}. In order to obtain reliable colours, we performed a crossmatch with the SDSS photometric catalogue, providing $g$- and $r$-band magnitudes for approximately 50\% of the sample. From this subsample, we plotted CFHT $r$-band magnitudes against SDSS $r$-band magnitudes to derive a calibration relation. For galaxies lacking SDSS photometry, absolute magnitudes were first computed from the calibrated apparent magnitudes (using the previously derived relation) and the galaxy redshift. Their $g-r$ colours were then estimated using the linear relations between $g-r$ and $r$ magnitude previously derived for early-type (E or S0) and late-type (S(B)a-S(B)d or Irr) galaxies according to their morphological classification. These estimated colours were subsequently used to compute stellar masses consistently. This approximation introduces an uncertainty of roughly 0.1~dex on the derived stellar masses. The mass-to-light ratio was then derived from the galaxy colour following \citet{Bell2003}, enabling the computation of stellar masses for all galaxies. Several diagnostic plots (e.g. $g-r$ colour versus stellar mass by morphological type, and the stellar mass distribution) were used to validate the calibration and mass estimation procedure, and are presented in Appendix~\ref{app:mass}.

In order to quantify the completeness of our spectroscopic sample, we selected six representative 1.2 deg$^2$ tiles across the two regions: three in each region, sampling diverse environments, namely cluster/group cores, dense filaments, and lower-density field regions. For each tile, galaxies were extracted from the SDSS photometric catalogue \citep{SDSSDR16} with $z_\mathrm{phot} < 0.03$ and visually inspected to remove extended stars and other contaminants. We also included potential galaxies with $0.03 < z_\mathrm{phot} < 0.06$ that were not point sources, to account for background objects that could be missing from our spectroscopic sample, following the methodology of \cite{EROPerseusOverview}. Stellar masses for these missing galaxies were computed in the same manner as for the main catalogue.

The resulting catalogue of missing galaxies allows us to compute the completeness fraction as a function of stellar mass. The completeness curve was fitted using a logistic function \citep[e.g.,][]{Press2007} of the form
\begin{equation}
f_\mathrm{comp}(M_\ast) = \frac{1}{1 + \exp[-k(\log_{10} M_\ast - \log_{10} M_{50})]},
\end{equation}
where $M_{50} = 10^{8.86}\,M_\odot$ is the mass at which the sample reaches 50\% completeness, and $k = 2.39$ determines the steepness of the transition. From this fit, our spectroscopic sample reaches 90\% completeness at $\log_{10}(M_\ast/M_\odot) \simeq 9.78$, while the average completeness over the range $7.5 \le \log_{10}(M_\ast/M_\odot) \le 12$ is approximately 69\%.

\begin{figure}
\centering
\includegraphics[width=0.35\textwidth]{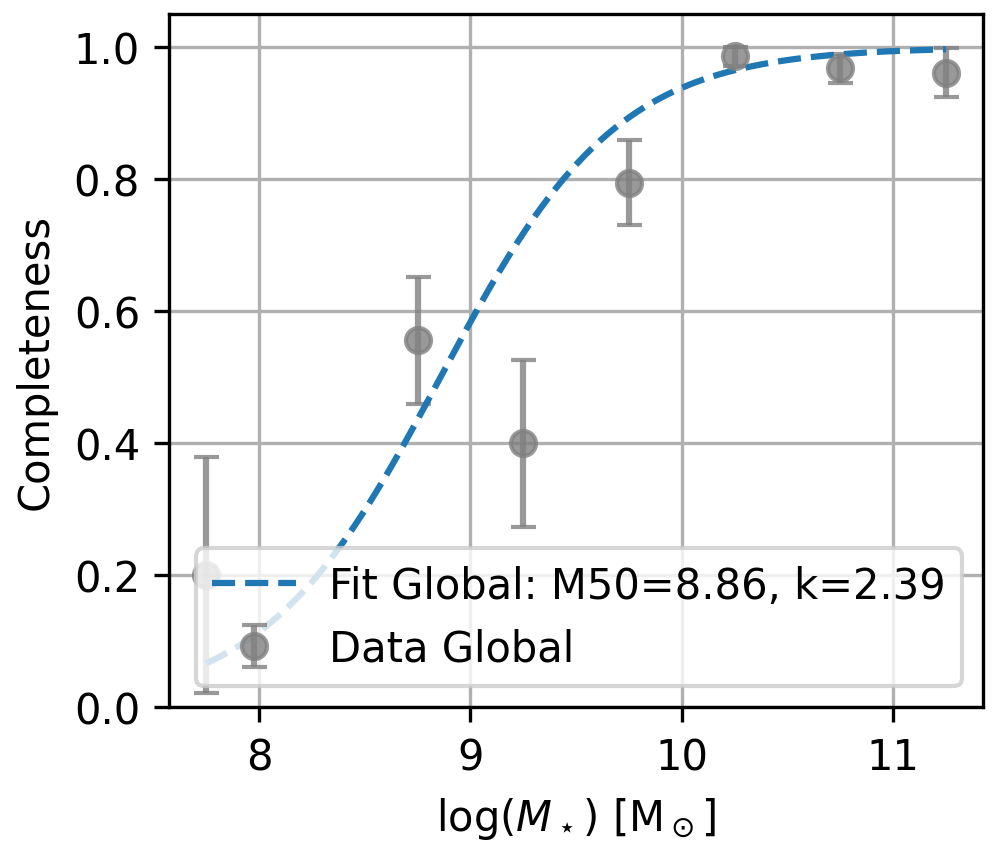}
\caption{Stellar mass completeness of the spectroscopic sample. Grey dots indicate the fraction of galaxies recovered in each stellar mass bin compared to the SDSS photometric reference \citep{SDSSDR16}, while the blue curve shows the logistic fit. The 50\% completeness threshold is at $\log_{10}(M_\ast/M_\odot) = 8.86$.}
\label{fig:mass_completeness}
\end{figure}

The completeness characterisation will be used in Sect.~\ref{sc:Results} to correct for mass-dependent selection effects when analysing the variation of galaxy properties with respect to filamentary and cluster environments.

\section{Results}\label{sc:Results}

\subsection{Properties of the PPSC}\label{sec:filament_size}

We present here the structural and physical properties of the two main regions of the PPSC.

\subsubsection{Filament Size: Length and Radius}

The main spine of the filaments in region~A extends over $\sim$48~Mpc, while the total filament length in that region sums to $\sim$127~Mpc. Region~B is more complex: its central skeleton stretches nearly 50~Mpc from A\,347 to the NGC\,383 group (Pisces cluster), two bifurcations near NGC\,383 each exceed 60~Mpc, and a branch from A\,262 extends $\sim$80~Mpc southward. Including smaller branches, the total filamentary length in region~B approaches $\sim$600~Mpc and occupies a box of order $3.1\times10^{5}$~Mpc$^{3}$.

\begin{figure}
\includegraphics[width=\columnwidth]{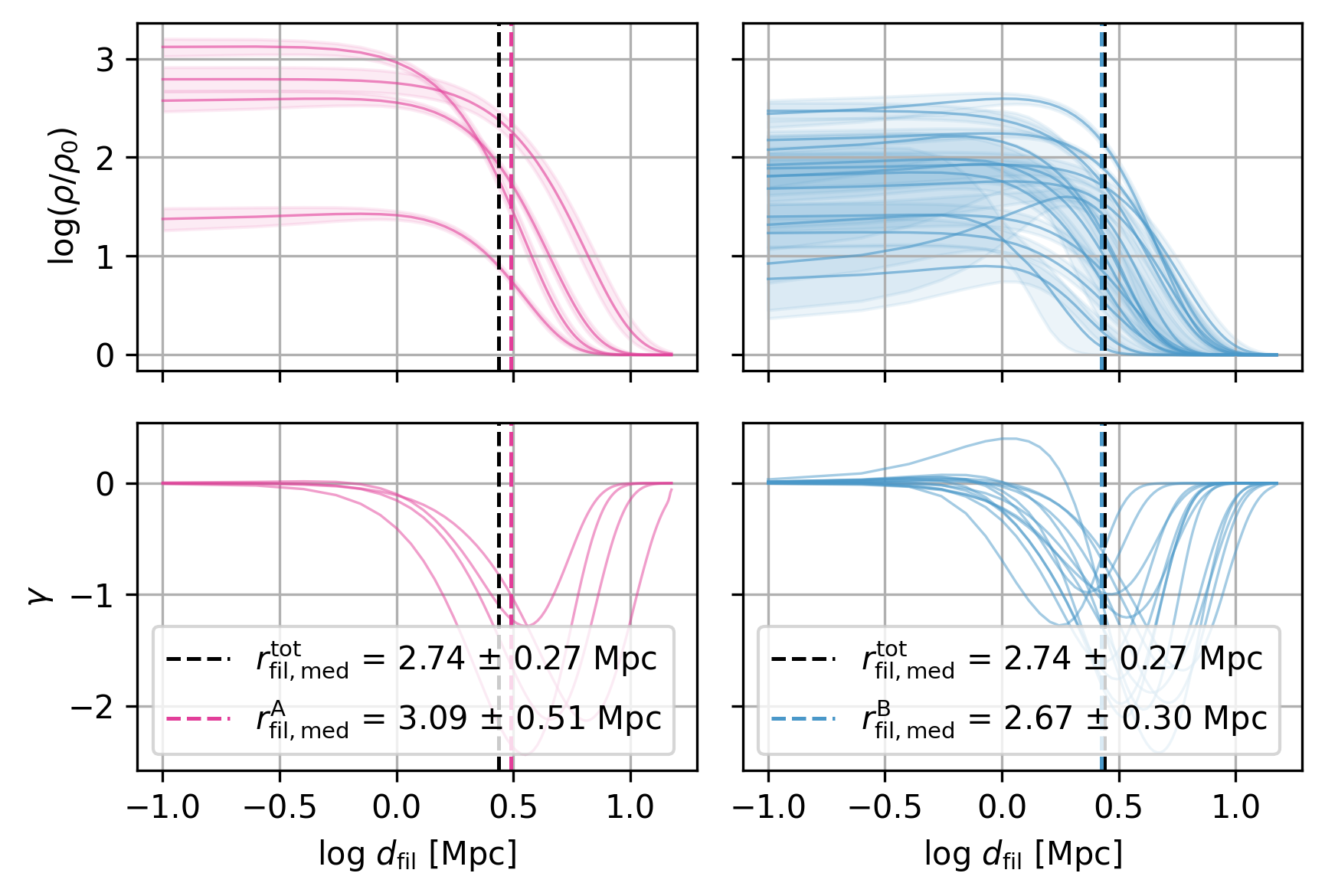}
\caption{Relative galaxy number density profiles, $\rho/\rho_0$, versus perpendicular distance to the filament spine $d_{\mathrm{fil}}$. Blue lines: region A; pink lines: region B. Solid lines show fitted profiles and shaded bands the 16–84\% MCMC percentiles. Dashed vertical lines mark the median $r_{\mathrm{fil}}$ for each region. Bottom panels: logarithmic slope $\gamma=d\log_{10}\rho/d\log_{10} d_{\mathrm{fil}}$; the minimum of $\gamma$ defines $r_{\mathrm{fil}}$.}
\label{fig:size_filament}
\end{figure}

After characterizing the filament lengths in regions~A and~B, we now examine their transverse extent. We estimate the filament radius $r_{\mathrm{fil}}$ from the projected galaxy distribution around each filament spine, using galaxies with $M_\star > 10^{8.5}\,M_\odot$. Radial number-density profiles are computed in logarithmic annuli and fitted with a half-Gaussian model following \citet{GalarragaEspinosa2022}; the characteristic radius is defined as the position of the minimum of the logarithmic slope of the fitted profile (see Appendix~\ref{app:filprofiles}). From this procedure, we obtain a median $r_{\mathrm{fil}} = 2.74 \pm 0.27$~Mpc. Region-specific values are $r_{\mathrm{fil,A}} = 3.09 \pm 0.51$~Mpc and $r_{\mathrm{fil,B}} = 2.67 \pm 0.30$~Mpc, i.e. filaments in region~A are $\sim$16\% larger, although this difference remains modest given the uncertainties. The offset could be partly explained by residual FoG effects from large Abell clusters, which tend to inflate measured radii in redshift space. The individual profiles, model fits, slope profiles and resulting radii are displayed in Fig.~\ref{fig:size_filament}.

Although our selection threshold is $M_\star > 10^{8.5}\,M_\odot$, the catalogue is fully complete only above $\sim10^{9.8}\,M_\odot$. As a consequence, our profiles are dominated by massive galaxies tracing the densest parts of the filaments, while fainter galaxies, which would better sample the diffuse envelope, are underrepresented. This limitation likely could lead to a slight overestimate of the filament width.

In observational studies, filament widths are typically inferred from projected galaxy-density profiles around filament spines identified in redshift surveys. \citet{SantiagoBautista2020}, for instance, analysed SDSS filaments over $0.02 < z < 0.15$ and reported typical radii of $\sim 2.5\,h_{70}^{-1}$~Mpc. They found that the galaxy number density contrast reaches $\simeq 3$ at $\sim3\,h_{70}^{-1}$~Mpc from the filament spine. 

For the PPSC, we measure the mean density contrast in number at a fixed distance of $3$~Mpc from the filament spines, finding $\delta_n \simeq 2.7$ across all filaments in regions~A and~B. This value is remarkably close to the SDSS measurements  at slightly higher redshifts, indicating that the PPSC filaments represent similarly dense, mature structures. Other surveys adopting similar contrast-based definitions \citep[e.g.][]{Alpaslan2015, Chen2017} find widths between 2 and 4~Mpc, depending on tracer selection, redshift, and smoothing scale. Our values therefore lie within the upper range of observational estimates, which is reasonable given the limited completeness of our sample.

In cosmological simulations, the full three-dimensional density field allows a more direct measurement of the matter distribution around filament spines. Early works such as \citet{Colberg2005} identified filaments between cluster pairs and defined their edge as the radius where the matter-density contrast falls below a fixed threshold ($\delta\rho/\rho \sim 1$), typically yielding radii of 1--2~Mpc. More recent analyses using skeleton-based approaches \citep[e.g.][]{Cautun2014, Kraljic2018, GalarragaEspinosa2020, Wang2024} fit analytical forms such as Gaussian or exponential functions to the radial density profiles of dark-matter particles or halos, finding similar values of $\sim$1~Mpc at $z\sim0$. The larger radii measured in our work are thus consistent with the trend that observationally derived widths, affected by projection, redshift distortions, and tracer selection, tend to exceed the intrinsic matter widths found in simulations.

Assuming a constant radius along each filament, the network occupies a total volume of $\sim(2.28\pm0.14)\times10^{4}$~Mpc$^3$, which forms the basis for our subsequent mass-density estimates.

\subsubsection{Mass density}

In order to estimate the mass density of the PPSC, we assign galaxies to clusters, filaments, or outskirts based on their distance to the nearest cluster or filament. The adopted boundaries are $r_\mathrm{200}$ for each group or cluster and $r_\mathrm{fil}$ for each filament. Stellar masses are corrected for incompleteness and converted to total halo mass assuming a stellar-to-halo mass ratio of $f_\star = 0.015$.

The mean matter density of the PPSC is $\rho_\mathrm{total} \simeq (9.4\pm0.5) \times 10^{10}~M_\odot\,\mathrm{Mpc}^{-3}$ over a total volume of $(6.37\pm0.14)\times10^4~\mathrm{Mpc}^3$. For comparison, \citet{Bohringer2021} report $\rho_\mathrm{Bohringer} \sim 8.6\times10^{10}~M_\odot\,\mathrm{Mpc}^{-3}$, in reasonable agreement given methodological differences. 

For further context, the cosmic mean matter density is $\rho_\mathrm{m,0} \simeq 2.8 \times 10^{10}~M_\odot\,\mathrm{Mpc}^{-3}$ \citep{Planck2018}. Thus, the PPSC exhibits a mean overdensity of $\delta \equiv \rho_\mathrm{PPSC}/\rho_\mathrm{m,0} - 1 \sim 2.4$, confirming that it represents a prominent supercluster-scale overdensity in the local Universe. Typical filament densities in the local Universe range from $5\times10^{10}$ to $10^{11}~M_\odot\,\mathrm{Mpc}^{-3}$ \citep{Kraljic2018, GalarragaEspinosa2022}, consistent with the densities measured here. 

The densities inferred for the PPSC thus fall squarely within the expected range for the most massive filamentary structures in the nearby Universe.

\subsection{Morphological segregation}

Understanding the spatial distribution of galaxy morphologies in relation to large-scale structures provides important insights into the environmental processes driving galaxy evolution. In this section, we investigate how morphological types segregate within the PPSC, with a particular focus on the contrast between the region A and B.

Table~\ref{tab:morphoreg} presents the distribution of morphological types and their median Sérsic indices in each environment. As anticipated, a systematic decline in the median Sérsic index is observed from elliptical to irregular galaxies, reflecting the structural diversity of these systems, from centrally concentrated spheroids to more diffuse and clumpy morphologies. This trend confirms the robustness of our morphological classification scheme.

A notable feature emerges for the irregular population in the region A, which exhibits a slightly elevated median Sérsic index ($n = 1.21 \pm 0.09$) compared to that observed in the region B ($n = 1.05 \pm 0.04$). This discrepancy may arise from surface brightness profiles affected by compact central components such as bulges or nuclear star-forming regions, leading to steeper inner light distributions. In dense cluster environments, additional mechanisms such as tidal stirring or galaxy harassment could also play a role in reshaping the outer structure of irregular galaxies \citep{Mastropietro2005}. Conversely, the irregular population in the region B constitutes a larger and more spatially extended component, typically located along the filamentary network or toward its outskirts, where galaxies undergo different types of interactions, experience slightly lower interaction rates, and evolve in relative isolation.

\begin{table}
\centering
\begin{tabular}{l l c c}
\hline\hline
\noalign{\vskip 1pt}
Region & Morphology & Count (\%) & Median Sérsic index \\
\hline
A 
& E   & 165 (19.0\%) & $3.18 \pm 0.12$ \\
& S0  & 239 (27.5\%) & $2.36 \pm 0.09$ \\
& S(B)a/S(B)d & 229 (26.4\%) & $1.24 \pm 0.09$ \\
& Irr & 139 (16.0\%)  & $1.21 \pm 0.09$ \\
\hline
B  
& E   & 51  (2.9\%)  & $3.89 \pm 0.21$ \\
& S0  & 176 10.1\%)  & $2.29 \pm 0.10$ \\
& S(B)a/S(B)d & 887 (50.9\%) & $1.05 \pm 0.04$ \\
& Irr & 404 (23.2\%) & $1.03 \pm 0.05$ \\
\hline
\end{tabular}
\caption{Morphological classification with counts and percentages (relative to the total number of galaxies in each region) and median Sérsic index by environment.}
\label{tab:morphoreg}
\end{table}

Table~\ref{tab:morphoreg} clearly reveals a morphological segregation: late-type galaxies (spirals and irregulars) are predominant in region B, while early-type galaxies (ellipticals and S0s) dominate region A. Indeed, the fraction of early-type systems is significantly higher in region A than in region B.

This difference reflects the contrasting environmental conditions and evolutionary stages of the two regions. The region B is more diffuse and dynamically younger, favouring the presence of gas-rich, star-forming galaxies with extended and irregular light profiles. In contrast, the region A corresponds to a more evolved, high-density environment in which frequent interactions and environmental quenching processes have transformed galaxies into early-type systems with more concentrated stellar distributions. This is consistent with our finding that filaments in the region A have smaller characteristic radii, likely due to the dominance of massive galaxies which tend to reduce the apparent filament width \citep{GalarragaEspinosa2020}.

\begin{figure*}
\centering
\includegraphics[width=1.\textwidth]{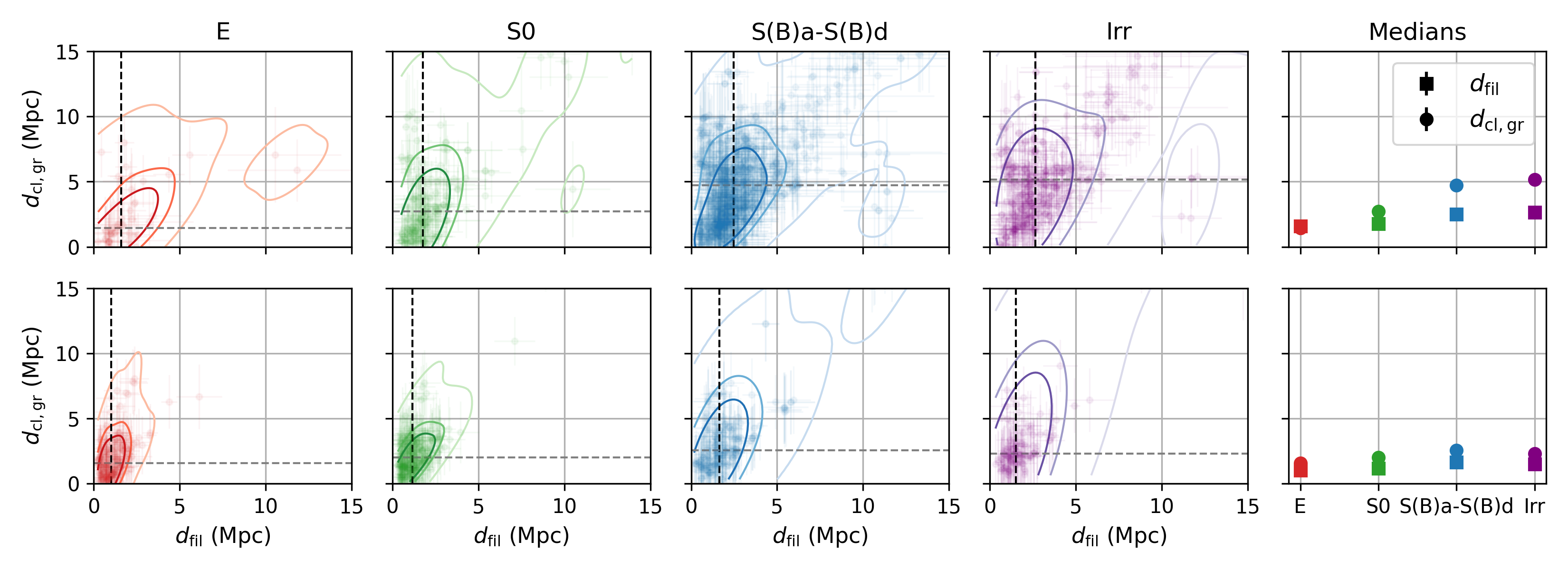}
\caption{
Distance to the filament spine $d_{\mathrm{fil}}$ versus distance to the nearest cluster $d_{\mathrm{cl}}$ for galaxies in two regions, grouped by morphological type. Top panels: region B. Bottom panels: region A. Each panel displays three Gaussian density contours corresponding to percentiles 0.5, 0.683, and 0.954. The rightmost panels summarise the median values of $d_{\mathrm{fil}}$ and $d_{\mathrm{cl}}$ for each morphological type.
}
\label{fig:combined_distance_filament_cluster}
\end{figure*}

Further insight into the environmental dependence of morphology is provided by Fig.~\ref{fig:combined_distance_filament_cluster}, which display the distance to the filament spine $d_{\mathrm{fil}}$ as a function of the distance to the nearest cluster $d_{\mathrm{cl,gr}}$, separated by morphological type. Each panel shows the distribution for one morphological class (E, S0, S, Irr), along with Gaussian density contours representing the main percentiles. The rightmost panels summarise the median distances for each type.

A clear environmental gradient emerges: early-type galaxies are preferentially located closer to clusters, and to a lesser extent, to filamentary structures. Late-type galaxies, by contrast, are typically found at greater distances from clusters and are more diffusely distributed with respect to the filament spine. This segregation is apparent in both the region B and region As, though it is more pronounced in the latter. These results are consistent with previous studies \citep[e.g.][]{Kraljic2017, Malavasi2017}, which have demonstrated that both galaxy morphology and star formation activity are strongly modulated by location within the cosmic web. In particular, early-type galaxies are preferentially found in dense environments such as nodes and cluster cores, while late-type systems dominate in filaments and less dense regions. Our analysis confirms this picture and further refines it within the context of the PPSC, highlighting the distinct evolutionary pathways of galaxies across its various substructures.

\begin{figure*}
\centering
\includegraphics[width=0.8\textwidth]{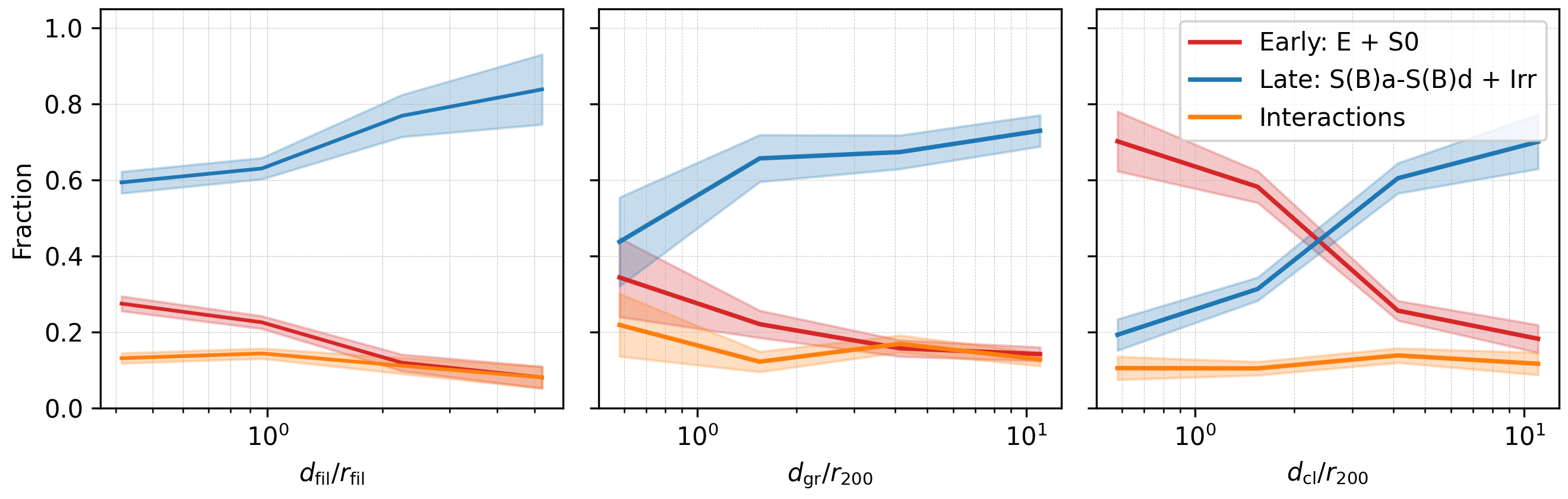}
\caption{Morphological fractions of galaxies as a function of normalised distances to the filament ($d_\mathrm{fil}/r_\mathrm{fil}$, left), to the nearest group centre ($d_\mathrm{gr}/r_\mathrm{200}$, middle), and to the nearest cluster centre ($d_\mathrm{cl}/r_\mathrm{200}$, right). Early-type (E + S0) galaxies are shown in red, late-type galaxies (S(B)a–S(B)d + Irr) in blue, while interacting systems (merger, tidal tail, ram-pressure stripping, shell, asymmetry in halo) are shown in orange. Shaded areas represent binomial uncertainties.}
\label{fig:TypeFractions_all}
\end{figure*}

Figure~\ref{fig:TypeFractions_all} presents the morphological fractions of galaxies as a function of normalised distances to the nearest cluster, group, and filament environments. 
When examining the distance to the filament spine ($d_{\mathrm{fil}}/r_{\mathrm{fil}}$), excluding galaxies within $r_{200}$ of clusters or groups, late-type systems dominate at all distances. Their fraction increases steadily outward, from $\sim$0.6 near the filament core to $\gtrsim$0.75 in the outskirts, suggesting that filaments are gas-rich environments hosting primarily star-forming galaxies, with only a modest early-type contribution near their spines. When considering the cluster-centric distance ($d_{\mathrm{cl}}/r_{\mathrm{200}}$), we observe the expected strong segregation: early-type galaxies dominate the dense cores, while late-types become increasingly common beyond the virial radius. The crossover between early- and late-type fractions occurs at $\sim2$–$3\,r_{\mathrm{200}}$, consistent with a transition from quiescent, morphologically transformed populations in cluster centres to more star-forming systems in the outskirts. In rich groups, the central fractions of early- and late-type galaxies are comparable, but late-types gradually dominate with increasing $d_{\mathrm{gr}}/r_{\mathrm{200}}$. This smoother trend reflects the milder environmental processing at intermediate densities, where tidal interactions and minor mergers remain efficient while hydrodynamical quenching is less pronounced.  

Overall, these three complementary trends highlight the progressive transition in galaxy populations across environments, from early-type, dominated, dynamically evolved cluster cores to late-type–rich filaments tracing the large-scale structure of the local cosmic web.

\subsection{Diversity of Local Interactions}

We now focus on the signs of galaxy interactions in our sample. Table~\ref{tab:interactions} summarizes the frequency of each interaction type observed in regions A and B. Overall, the total fraction of strongly interacting galaxies is slightly higher in the region B (12.02\%) than in the region A (10.6\%), indicating that filaments host a comparatively larger population of perturbed galaxies. This is consistent with the picture of enhanced gravitational and environmental effects along large-scale structures.

\begin{table}[h!]
\centering
\begin{tabular}{l l c }
\hline\hline
\noalign{\vskip 1pt}
Region & Type of interaction & Count (\%) \\
\hline
A
& Merger & 14 (1.6\%) \\
& Tidal tail & 11 (1.3\%) \\
& Ram pressure stripping & 2 (0.2\%) \\
& Shell & 1 (0.1\%) \\
& Asymmetry in halo & 68 (7.8 \%) \\
\hline
B
& Merger & 26 (1.5\%) \\
& Tidal tail & 27 (1.76\%) \\
& Ram pressure stripping & 0 (0.0\%) \\
& Shell & 4 (0.2\%) \\
& Asymmetry in halo & 166 (9.5\%) \\
\hline
\end{tabular}
\caption{Number and percentage of galaxies displaying strong interaction features in the regions A and B. Percentages are relative to the total galaxy counts in each respective region (A: 10.6\%; B: 12.02\%).}
\label{tab:interactions}
\end{table}

Many clusters and groups reside at the nodes of filaments, including the Perseus Cluster, the Pisces Cluster, and Abell 262. Their central galaxies are typically cD, such as NGC 1275 in Perseus cluster\citep{Salome.2008}. Beyond these massive structures, numerous galaxies exhibit clear signs of local interaction: minor and major mergers are observed along filaments, while others display tidal tails or shells, classical tracers of past or ongoing interactions \citep{Prieur.1988, Teuben.2023} (see Fig.~\ref{interaction_examples}).

We classify galaxies by interaction type and analyze the fraction of interacting galaxies as a function of both the distance to the filament spine ($d_{\mathrm{fil}}$) and to the nearest cluster ($d_{\mathrm{gr,cl}}$). 

Figure~\ref{fig:TypeFractions_all} shows also the fraction of interacting galaxies as a function of $d_{\mathrm{cl,gr}}/r_{\mathrm{cl}}$ for both groups and clusters. A clear difference emerges between the two environments. For groups, the interaction fraction steadily decreases with increasing distance from the centre, consistent with the idea that gravitational perturbations and mergers are favoured in their dense cores, where galaxy densities are relatively high and velocity dispersions are low. In contrast, for clusters, the fraction of interacting systems follows the behaviour described previously: it rises from the centre toward intermediate radii, peaking near the virial boundary ($\sim r_{\mathrm{cl}}$), before declining again in the outskirts. 

This difference reflects the strong dependence of gravitational interactions on environmental conditions. As shown in previous studies \citep{DeLucia2006,Boselli2006}, optical signatures of interactions are most sensitive to tidal perturbations, which are efficient in high-density but low-velocity environments such as groups. In clusters, the high velocity dispersions tend to suppress mergers and strong tidal encounters, except in their peripheral regions where relative velocities are moderate and prolonged interactions can occur \citep{Zhang2008}.

Overall, these findings emphasize the crucial role of local gravitational interactions in shaping galaxy properties within large-scale structures, with filaments acting as intermediary zones that promote interactions before cluster accretion. Appendix~\ref{app:fraction_details} shows that the density of interactions closely follows the overall galaxy density as a function of distance from filaments and clusters, indicating that the observed decrease in interaction fraction is largely due to the lower galaxy density at larger distances rather than an intrinsic reduction in interaction probability.

The fractions of interacting galaxies observed in our sample can be compared with the recent statistics published by \citet{Sola2025}, who analyzed a large sample of nearby galaxies (19\,387 galaxies with $z \le 0.02$), primarily from the DES $r$-band data, particularly within the UNIONS project. In their study, \cite{Sola2025} distinguishes galaxies showing streams, mergers, or signs of minor interactions, finding an overall fraction of $\sim 11.9\%$ of galaxies exhibiting interaction features in the local field. This provides a valuable benchmark for comparison with our sample. 

Our results for both regions A and B fall close to this percentage, indicating overall consistency. However, a notable difference arises in the distribution of interaction types: whereas \cite{Sola2025} report a majority of fine streams and subtle minor interactions, our filaments show a higher fraction of halo asymmetries and a comparable number of mergers and tidal tails. This observation supports the view that filaments, as intermediate large-scale structures between clusters and the field, enhance both gravitational and environmental interactions. Moreover, by distinguishing interactions according to their location, we find that the overall interaction fraction slightly decreases with distance from the filament spine and cluster peripheries, highlighting the environment-dependent nature of galaxy interactions.

\subsection{Mass segregation and stellar mass function}

Building upon the morphological segregation discussed above, we now examine the stellar mass function (SMF) to trace how galaxy stellar masses vary across the PPSC and depend on environment. The global SMF, computed for the full sample regardless of the A/B regions, provides an integrated view of the stellar mass content within the structure and quantifies the relative contributions of early- and late-type systems.

\begin{figure}
\centering
\includegraphics[width=0.8\linewidth]{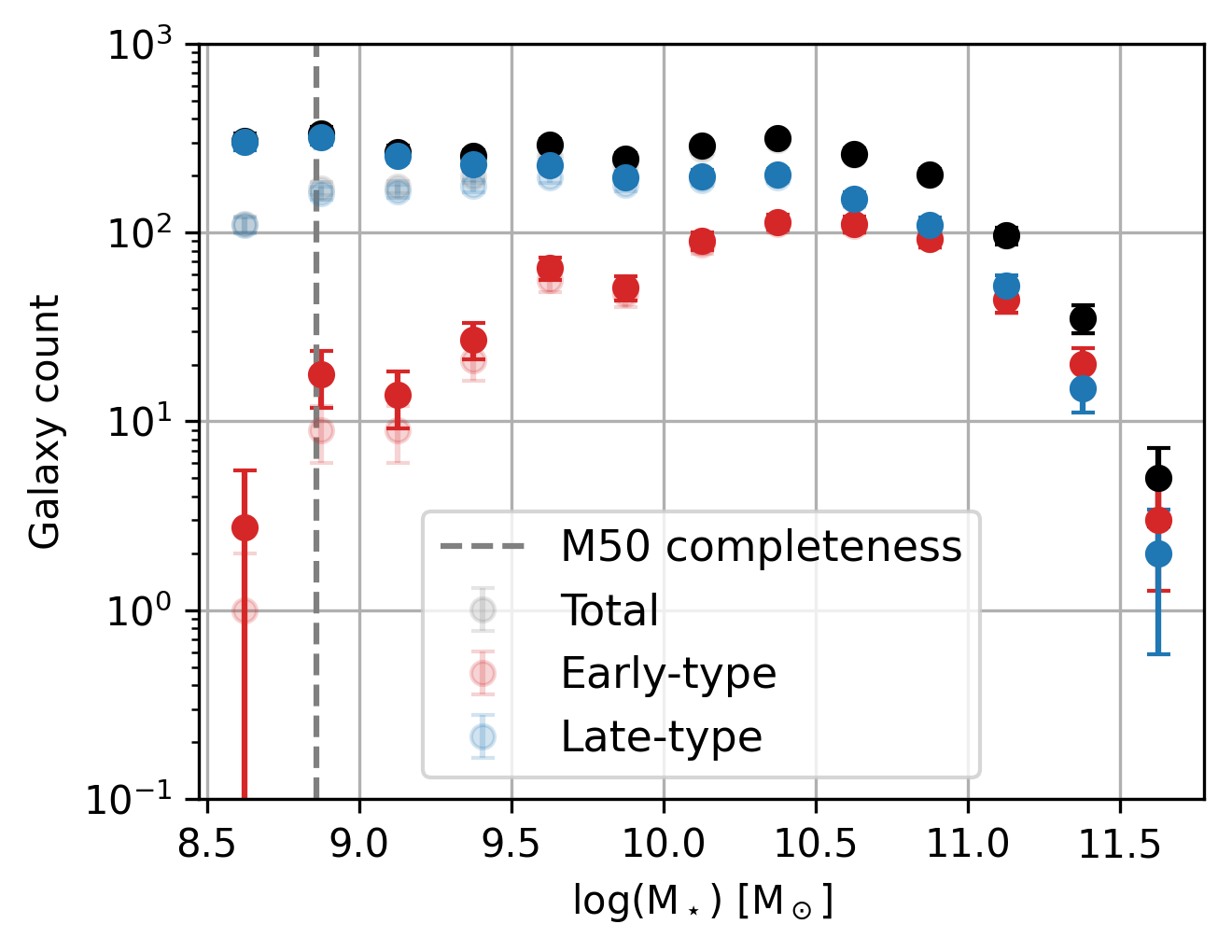}
\caption{
Stellar mass functions (SMFs) separated by morphological type. Early-type galaxies (E and S0) are shown in red, and late-type galaxies (S(B)a–S(B)d and Irr) in blue. Transparent symbols correspond to raw counts, while filled symbols include completeness corrections. Error bars denote Poisson uncertainties.
}
\label{fig:smf_morpho}
\end{figure}

Figure~\ref{fig:smf_morpho} shows the global SMF separated by morphology. Late-type galaxies (S(B)a-S(B)d/Irr; blue) dominate at low and intermediate masses, while early-type galaxies (E/S0; red) become dominant at the high-mass end, consistent with the expected morphology–mass relation.

\begin{figure*}
\centering
\includegraphics[width=1\linewidth]{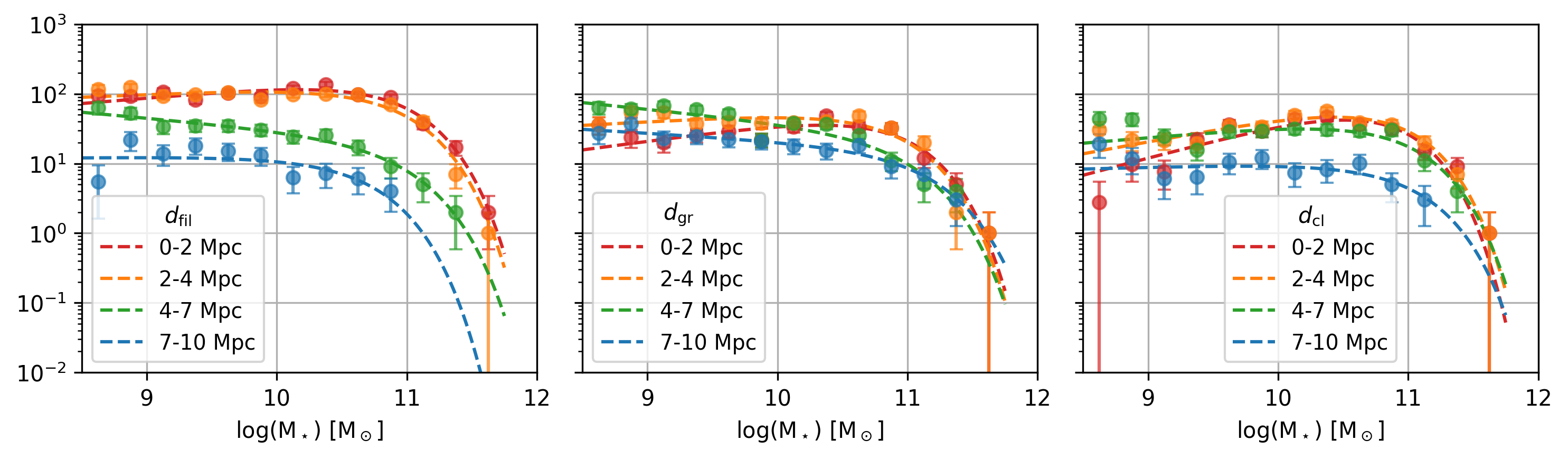}
\caption{
SMFs as a function of characteristic distance, based on completeness-corrected stellar masses. Distance bins are color-coded: 0–2\,Mpc (red), 2–4\,Mpc (orange), 4–7\,Mpc (green), and 7–10\,Mpc (blue).
\textit{Left}: Perpendicular distance to the filament axis ($d_{\mathrm{fil}}$), isolating galaxies outside groups and clusters.
\textit{Middle}: Projected distance from group centers ($d_{\mathrm{gr}}$).
\textit{Right}: Projected distance from cluster centers ($d_{\mathrm{cl}}$).
Best-fit Schechter parameters for each bin, with $1\sigma$ uncertainties, are shown in Table \ref{tab:smf_params}.}
\label{fig:smf_environment}

\end{figure*}

To probe environmental variations, Fig.~\ref{fig:smf_environment} presents the SMFs as a function of projected distance from the filament spine ($d_{\mathrm{fil}}$, top), from group centers ($d_{\mathrm{gr}}$, middle), and from cluster centers ($d_{\mathrm{cl}}$, bottom). Distance bins are color-coded as 0–2\,Mpc (red), 2–4\,Mpc (orange), 4–7\,Mpc (green), and 7–10\,Mpc (blue). Galaxies belonging to clusters or groups are excluded from the filament-centric profiles to isolate the diffuse and inter-filament populations.

The SMFs are fitted with a single Schechter function \citep{Schechter1976, EROPerseusOverview}:
\[
\Phi(M) = \phi^* \left(\frac{M}{M^*}\right)^\alpha \exp\left(-\frac{M}{M^*}\right),
\]
where $M^*$ marks the characteristic stellar mass (“knee”), $\alpha$ is the faint-end slope, and $\phi^*$ the normalization. Best-fit parameters are listed in Table~\ref{tab:smf_params}.

\begin{table}[htbp]
\centering
\begin{tabular}{lcccc}
\hline\hline
Bin (Mpc) & $\alpha$ & $\log(M_\star / M_\odot)$ & $\phi_\star$ (N dex$^{-1}$) \\
\hline
$d_{\mathrm{fil}}$ & & & \\
$[0,2]$   & $-0.87 \pm 0.04$ & $11.01 \pm 0.10$ & $74.8 \pm 10.0$ \\
$[2,4]$   & $-1.00 \pm 0.04$ & $11.15 \pm 0.12$ & $47.6 \pm 7.4$ \\
$[4,7]$   & $-1.24 \pm 0.04$ & $11.22 \pm 0.30$ & $6.2 \pm 2.2$ \\
$[7,10]$  & $-0.86 \pm 0.20$ & $10.30 \pm 0.32$ & $9.8 \pm 4.9$ \\
\hline
$d_{\mathrm{gr}}$ & & & \\
$[0,2]$   & $-0.82 \pm 0.08$ & $11.03 \pm 0.17$ & $25.5 \pm 5.8$ \\
$[2,4]$   & $-0.97 \pm 0.05$ & $11.14 \pm 0.17$ & $22.4 \pm 4.7$ \\
$[4,7]$   & $-1.13 \pm 0.06$ & $10.92 \pm 0.26$ & $14.8 \pm 5.1$ \\
$[7,10]$  & $-1.13 \pm 0.05$ & $11.24 \pm 0.25$ & $6.4 \pm 1.9$ \\
\hline
$d_{\mathrm{cl}}$ & & & \\
$[0,2]$   & $-0.70 \pm 0.06$ & $10.97 \pm 0.14$ & $32.4 \pm 6.4$ \\
$[2,4]$   & $-0.70 \pm 0.07$ & $10.92 \pm 0.11$ & $39.0 \pm 5.7$ \\
$[4,7]$   & $-1.03 \pm 0.08$ & $11.33 \pm 0.37$ & $12.4 \pm 5.0$ \\
$[7,10]$  & $-1.16 \pm 0.11$ & $11.79 \pm 2.02$ & $1.9 \pm 2.6$ \\
\hline
\end{tabular}
\caption{Best-fit Schechter parameters for the SMFs in different environments and distance bins within the PPSC. Distance bins correspond to separations from the filament spine ($d_{\mathrm{fil}}$), group centers ($d_{\mathrm{gr}}$), and cluster centers ($d_{\mathrm{cl}}$). Uncertainties represent $1\sigma$ errors from the fit.}
\label{tab:smf_params}
\end{table}

The faint-end slope $\alpha$ of the stellar mass function exhibits a clear environmental dependence. In filaments, $\alpha$ steepens from $-0.87 \pm 0.04$ in the inner bin (0–2 Mpc) to $-1.24 \pm 0.04$ at intermediate distances (4–7 Mpc), before slightly flattening in the outermost bin. This trend indicates that low-mass galaxies become increasingly abundant in more diffuse filamentary environments. Similarly, group environments show a steepening of $\alpha$ with distance, from $-0.82 \pm 0.08$ in the core to $-1.13 \pm 0.085$ in the outermost bin.

In contrast, cluster-centric SMFs reveal a remarkably flat faint-end slope in the innermost bin: $\alpha = -0.70 \pm 0.06$ at 0–2 Mpc. This is significantly shallower than typical values reported in other studies, including the Perseus cluster where $\alpha \sim -1$ \citep[e.g.,][]{Merluzzi2010}. Such a flat slope highlights the strong suppression of low-mass galaxies in dense cluster cores, consistent with efficient environmental quenching, tidal stripping, and ram-pressure removal of gas from satellite galaxies. At larger cluster-centric distances, $\alpha$ steepens gradually to $-1.16 \pm 0.11$, reflecting the increasing relative abundance of low-mass galaxies in lower-density outskirts.

These trends are consistent with previous studies both in the Perseus region \citep{EROPerseusOverview} and other cluster \citep[e.g.,][]{Merluzzi2010}. For instance, GAMA \citep{Sbaffoni2025} reports a systematic steepening of the faint-end slope in lower-density environments and filaments. A modest depression around $\log(M_\star/M_\odot) \sim 9.5$ is visible in some cluster SMFs, potentially marking a transitional population between quenched low-mass satellites and more massive galaxies; this feature is not explicitly modelled in our single-Schechter fits.

The characteristic stellar mass $M^*$ remains relatively stable across environments (typically $\log(M^*/M_\odot) \sim 10.8-11.0$), with a mild increase near the centers of clusters and groups, reflecting the survival and dominance of massive galaxies in the deepest potential wells. The normalization $\phi^*$ traces the local density contrast, being highest in cluster cores, intermediate in filaments, and lower in group outskirts.

Overall, these results reinforce the picture of environmental mass segregation in the PPSC: massive early-type galaxies dominate dense nodes, while low-mass, late-type galaxies preferentially populate filaments and the supercluster periphery. The exceptionally flat $\alpha$ in cluster cores quantitatively illustrates the strong suppression of low-mass satellites, providing a direct signature of the physical mechanisms—quenching, stripping, and pre-processing—that shape galaxy populations across the cosmic web.

\section{Discussion}\label{sc:Discussion}

Our study of the PPSC reveals a complex interplay between galaxy properties and their large-scale environments, highlighting the diversity of evolutionary pathways within a single cosmic structure. The contrast between region~A, dominated by the massive Perseus and AWM7 clusters, and region~B, which encompasses a network of groups and filamentary bridges, illustrates distinct stages of galaxy evolution driven by environmental mechanisms operating at different efficiencies and timescales. Region~A thus represents the dynamically mature end of the environmental sequence, where galaxies have already undergone substantial environmental processing, while region~B traces younger, less virialized structures where galaxies are still actively assembling and transforming.

Beyond this large-scale dichotomy, we refined the environmental classification of the PPSC by distinguishing galaxies located (i) within $R_{200}$ of clusters, (ii) within $R_{200}$ of rich groups, (iii) along the filamentary skeleton, and (iv) in the outskirts. The adopted $R_{200}$ values were extracted from heterogeneous literature sources and should therefore be considered as indicative rather than exact limits. Nevertheless, they provide a useful estimate of the virial extent of the main halos. For the filaments, several radius measurement methods exist in the literature, each with specific biases and sensitivities to scale definition \citep[e.g.,][]{Colberg2005, Alpaslan2015, Cautun2014, Chen2017, Kraljic2018, GalarragaEspinosa2020, SantiagoBautista2020}. In our case, the median filament radius of $r_{\mathrm{fil}} \simeq 2.7$~Mpc derived from galaxy-density profiles (see Sect.~\ref{sec:filament_size}) is fully consistent with observational measurements in the local Universe up to $z \lesssim 0.1$ \citep[e.g.,][]{Alpaslan2015, Chen2017, SantiagoBautista2020}. This confirms that the Perseus–Pisces filaments share structural and density properties typical of mature, well-defined cosmic-web filaments, providing a robust framework for our subsequent environmental analysis.

\begin{figure}
\centering
\includegraphics[width=0.4\textwidth]{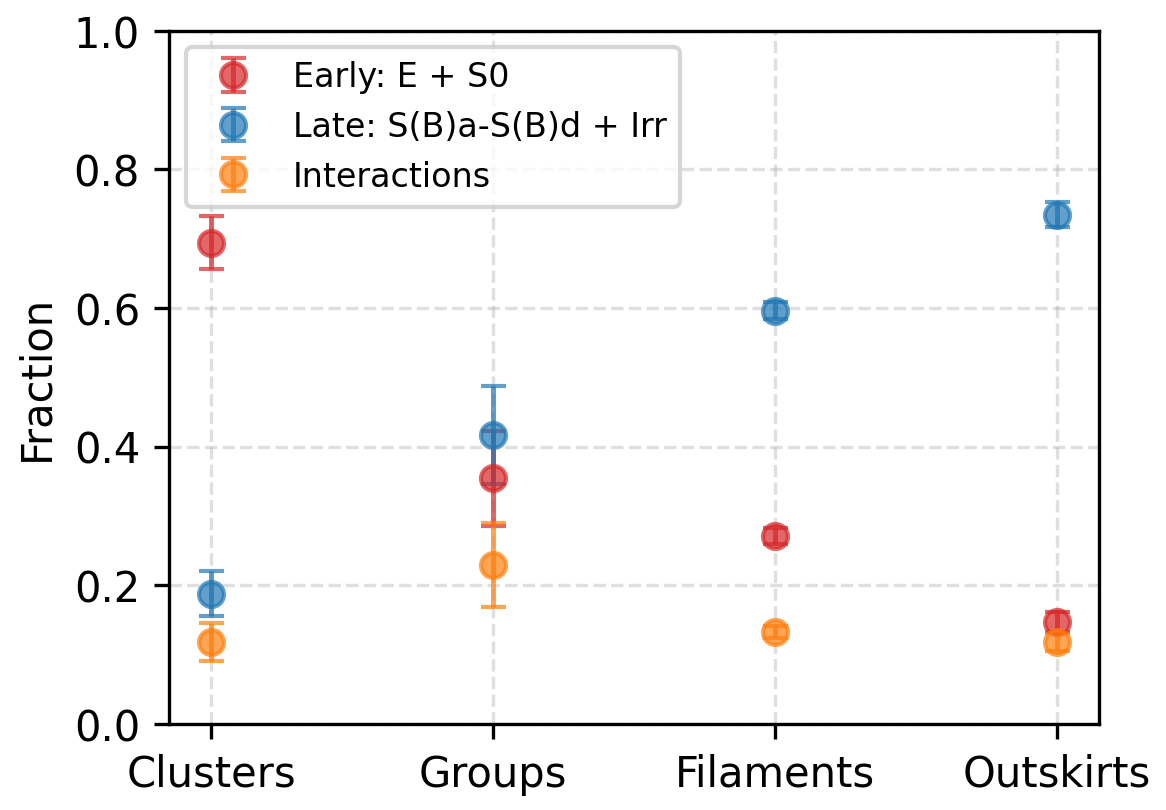}
\caption{Total type fraction of galaxies in different kinds of environments: clusters, groups, filaments, and outskirts. Early-type (E + S0) galaxies are shown in red, late-type galaxies (S(B)a–S(B)d + Irr) in blue, while interacting systems (merger, tidal tail, ram pressure stripping, shell, asymmetry in halo) are shown in orange. Error bars correspond to Poisson uncertainties.}
\label{fig:totalfraction}
\end{figure}

Figure~\ref{fig:totalfraction} summarizes the global morphological composition of galaxies across these environments. A clear gradient emerges: the fraction of early-type systems decreases steadily from the cluster cores toward the filaments and outskirts, while late-type galaxies become increasingly prevalent. This behaviour reflects the expected decline of environmental density and interaction frequency along the cosmic web.

A comparison with the Virgo filaments studied by \citet{Castignani2022} reinforces this picture. Although their classification scheme does not explicitly separate interacting galaxies from morphological types, potentially altering direct fraction comparisons, the overall trends are consistent. They show a continuous increase of late-type dominance with decreasing environmental density. In the PPSC, we find similar relative behaviour with a late-type fraction in filaments of $\sim 0.6$, consistent within uncertainties, while the clusters show a lower late-type fraction ($\sim 0.2$). However, when including galaxies identified as interacting, many of which display morphological asymmetries likely induced by ram-pressure stripping, the effective late-type fraction in clusters rises to $\sim0.4$, yielding values broadly compatible with the Virgo results. For rich groups, our  late-type fraction around $0.42$ remains slightly below the Virgo value but again becomes consistent ($\sim0.55$) when accounting for the interacting systems. In the outskirts, the late-type fraction approaches $\sim0.75$, in agreement with the high prevalence of unperturbed, star-forming galaxies found in low-density environments.

Turning to the interacting population, we find that the overall fraction of systems showing signs of mergers, tidal distortions, or asymmetries decreases from about 25\% in groups to 10\% in the outskirts. In clusters, the fraction is lower ($\sim$12\%) than in groups, which can be attributed to the high velocity dispersions that inhibit slow mergers and major encounters \citep[e.g.,][]{DeLucia2006, Boselli2006}. Instead, cluster galaxies experience hydrodynamical and gravitational processes, like ram-pressure stripping, tidal harassment, and strangulation, that effectively quench star formation and reshape morphologies without necessarily producing strong interaction features \citep[e.g.,][]{Gunn1972, Moore1996, Boselli2006, Peng2015}. Within clusters, we also find that the fraction of interacting galaxies remains roughly constant with distance from the centre, suggesting that these mechanisms operate efficiently throughout the virialized region.

Outside clusters, interactions are driven by different processes. In groups and filaments, where relative velocities are lower and local densities moderate, gravitational encounters and mergers occur more frequently. We observe a gradual decrease in the interaction rate with distance from the filament spine, consistent with findings by \citet{Moutard2022}. Moreover, as shown by \citet{Sola2025}, low surface brightness signatures trace the cumulative dynamical activity of galaxies in intermediate-density environments and highlight the long-lasting imprint of tidal perturbations and environmental quenching.

It is important to note that our results are limited by stellar-mass completeness. The analysis is robust above $\log(M_\star/M_\odot) \gtrsim 9.9$, but the incompleteness at lower masses likely underestimates the contribution of faint, gas-rich systems. Consequently, the fractions and evolutionary trends discussed here should be viewed as conservative estimates, potentially missing a population of low-mass, star-forming, or recently quenched galaxies.

The stellar-mass distribution itself reflects the underlying environmental hierarchy: clusters host the most massive, early-type systems, while groups and filaments exhibit broader and more gradual mass distributions, consistent with ongoing assembly and a higher diversity of evolutionary stages. The outskirts, in turn, contain predominantly low-mass, late-type galaxies that have experienced minimal environmental processing.

\section{Conclusion}

This study provides a comprehensive analysis of galaxy properties and their environmental dependencies within the Perseus–Pisces Supercluster, one of the most prominent overdensities in the nearby Universe. Using deep CFHT–MegaCam $r$-band imaging, we examined two main regions: the dynamically evolved region~A, dominated by the Perseus and AWM7 clusters, and the youngest region~B, which encompasses a complex network of groups and filamentary bridges extending after the Pisces cluster.

After correcting redshift distortions and reconstructing the cosmic web skeleton with the \texttt{DisPerSE} algorithm, we derived quantitative environmental metrics, the distances to the nearest cluster or group centre and to the filament spine. This allowed us to refine the environmental classification of galaxies into four categories: cluster, group members, filament galaxies, and outskirts galaxies. Although the $R_{200}$ values are gathered from heterogeneous sources, they provide a consistent first-order approximation of virial radii. For the filaments, we measured a typical radius of $r_{\mathrm{fil}} \sim 2.7 \pm 0.3$~Mpc, in line with local observational estimates and measurements up to $z \lesssim 0.1$ \citep[e.g.,][]{Alpaslan2015, Chen2017, SantiagoBautista2020}, confirming that the PPSC filaments share the geometric and physical characteristics of other nearby cosmic-web structures. However, the mean matter density of the PPSC, $\rho_\mathrm{total} \simeq (9.4\pm0.5)\times10^{10}~M_\odot\,\mathrm{Mpc}^{-3}$, indicates that these filaments are relatively dense compared to typical local filaments, consistent with observed ranges of $5\times10^{10}$–$10^{11}~M_\odot\,\mathrm{Mpc}^{-3}$ \citep{Kraljic2018, GalarragaEspinosa2022}.

Our morphological and stellar-mass analyses reveal clear environmental trends, with early-type galaxies dominating the densest regions and late-type fractions increasing toward lower-density environments such as groups, filaments, and outskirts. These trends are broadly consistent with other studies of nearby filamentary structures, such as \citet{Castignani2022} for the Virgo filaments, supporting the view that filaments act as transitional environments where galaxies experience milder quenching and continued growth before entering cluster cores.
Interactions also play a role in galaxy evolution across the PPSC, although their nature varies with environment. In clusters, high velocities suppress slow mergers, so hydrodynamical processes dominate, whereas in groups and filaments, tidal encounters and minor mergers contribute to moderate pre-processing \citep[e.g.,][]{Hidding2016,Moutard2022, Sola2025}. The stellar mass distribution reinforces this picture: clusters host the most massive galaxies, while groups and filaments contain a broader mass range with numerous low-mass systems. This contrast highlights two evolutionary stages within the PPSC, from virialized, quiescent cluster cores to actively assembling, star-forming filamentary environments.

Although our results provide a detailed snapshot of galaxy evolution within the PPSC, limitations remain. The stellar-mass completeness threshold ($M_\star \gtrsim 10^{9.8}\,M_\odot$) restricts the analysis of the lowest-mass populations, which are crucial to understanding the earliest phases of pre-processing. Furthermore, disentangling the relative influence of halo mass, local density, and filamentary geometry will require deeper spectroscopic coverage and multi-wavelength diagnostics of gas and star formation.

In summary, the Perseus–Pisces Supercluster exemplifies the complex, multi-scale interplay between galaxy mass, morphology, and large-scale environment. Clusters dominate through hydrodynamical and gravitational quenching, groups through tidal pre-processing and mergers, and filaments through gentle dynamical perturbations and continued accretion. Together, these processes shape the observed diversity of galaxy properties across the supercluster. Building upon the reconstructed cosmic web presented here, forthcoming analyses will investigate the alignment of galaxy spins and angular momentum with filamentary structures \citep[e.g.,][]{Welker2019,Castignani2022, Laigle2025}, further probing the imprint of the cosmic web on galaxy evolution.

\begin{acknowledgements}
We warmly thank the staff at the Canada–France–Hawaii Telescope for their dedication and support in obtaining the observations that made this work possible.
This research has made use of the NASA/IPAC Extragalactic Database (NED),
which is operated by the Jet Propulsion Laboratory, California Institute of Technology, under contract with the National Aeronautics and Space Administration. We also acknowledge the use of the HyperLEDA database, the tools and catalogs provided by the CDS, and the publicly available data from the 2MRS, SDSS, and FASHI surveys, which were essential for the analyses presented in this work.
\end{acknowledgements}

\bibliographystyle{aa} 
\bibliography{ref}

\begin{appendix}

\section{Skeleton construction: supplementary material}\label{app:skeleton}

This appendix provides additional details and figures related to the construction of the filamentary skeleton described in the main text.

\subsection{Correction of redshift-space distortions}

\begin{figure*}
\centering
\includegraphics[width=0.4\textwidth]{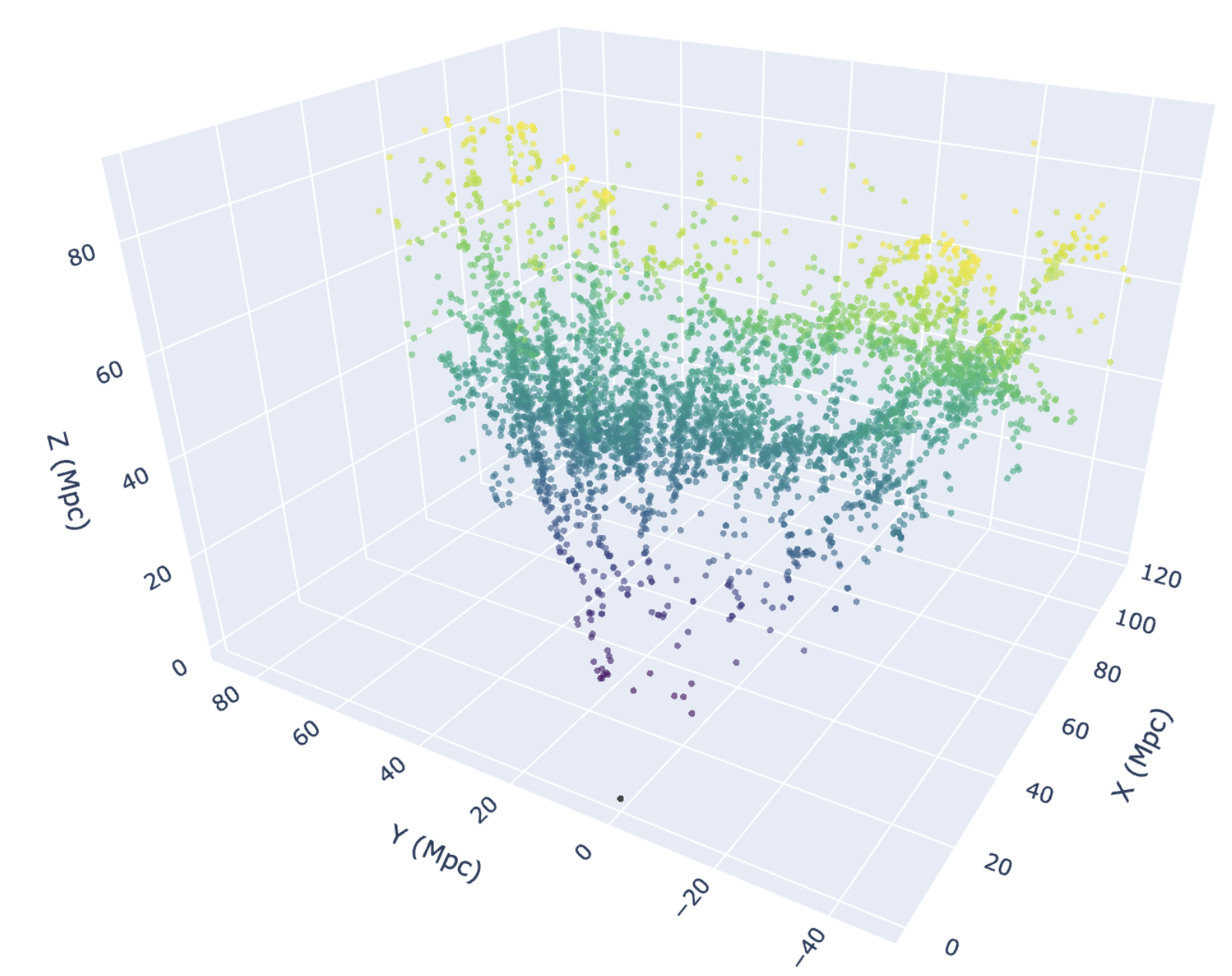}
$\hspace{0.1cm}$
\includegraphics[width=0.4\textwidth]{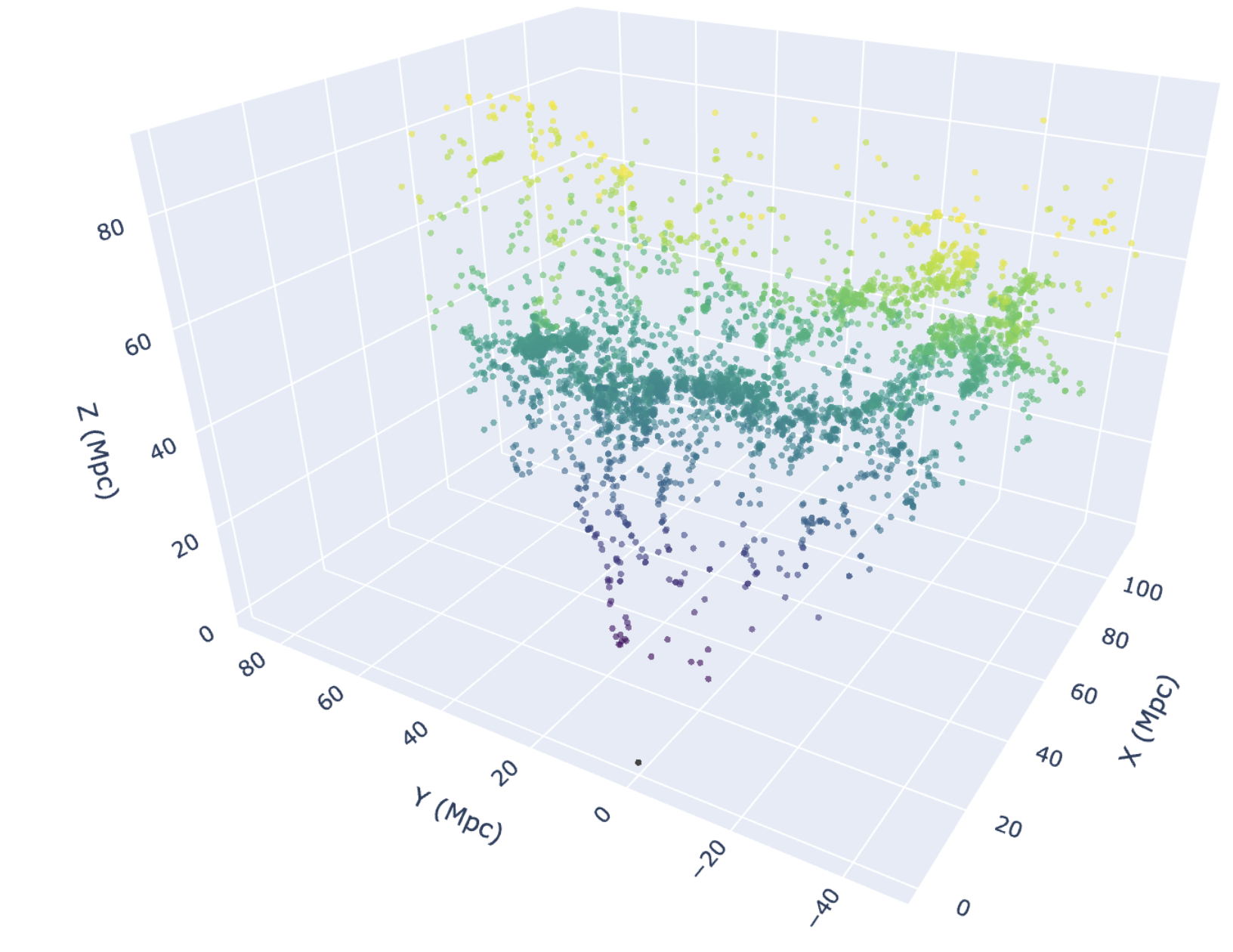}\smallskip\\
\caption{Redshift-space distortion correction. \textit{Left:} Galaxy distribution before FoF correction. \textit{Right:} After collapsing groups to their median redshift.}
\label{fig:FoG_correction}
\end{figure*}

The correction of redshift-space distortions is performed using a Friends-of-Friends (FoF) algorithm, with linking parameters $\Delta z \leq 0.005$ and angular separation $< 0.4^\circ$, and a minimum group richness of 10. Groups are collapsed to their median redshift positions. The effect of this correction is shown in Fig.~\ref{fig:FoG_correction}.

\subsection{Construction of the PPSC skeleton}

\begin{figure}
\centering
\includegraphics[width=0.4\textwidth]{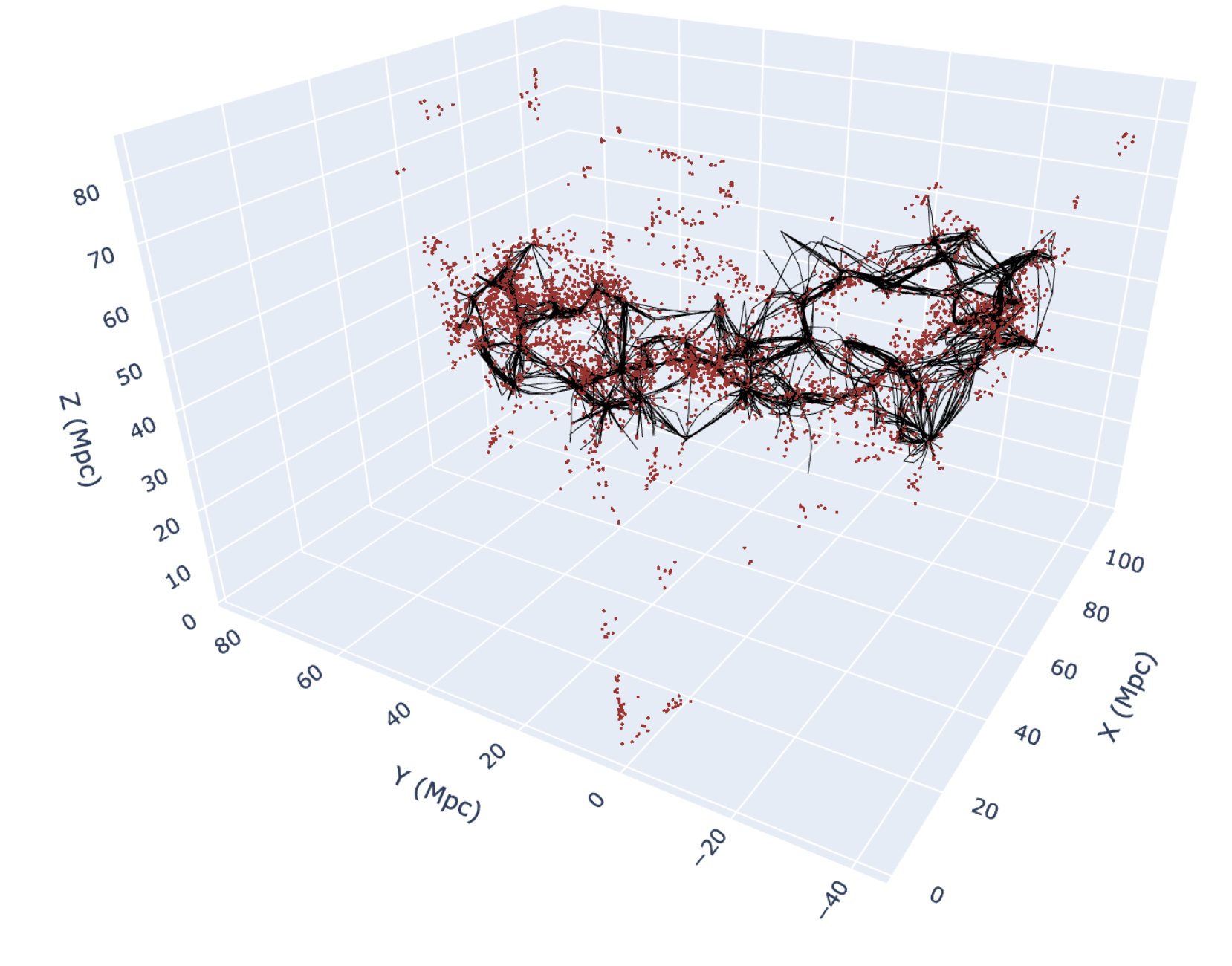}
\caption{Filaments extracted from 100 realisations of the galaxy distribution. Each segment is shown in black.}
\label{fig:all_realizations}
\end{figure}

The skeleton is derived from 100 realisations of the galaxy distribution, each obtained by randomly removing 5\% of the galaxies. For each realisation, \texttt{DisPerSE} is run with a $3\sigma$ persistence threshold and two smoothing iterations of the skeleton. The superposition of all resulting filaments is shown in Fig.~\ref{fig:all_realizations}.

\begin{figure}
\centering
\includegraphics[width=0.4\textwidth]{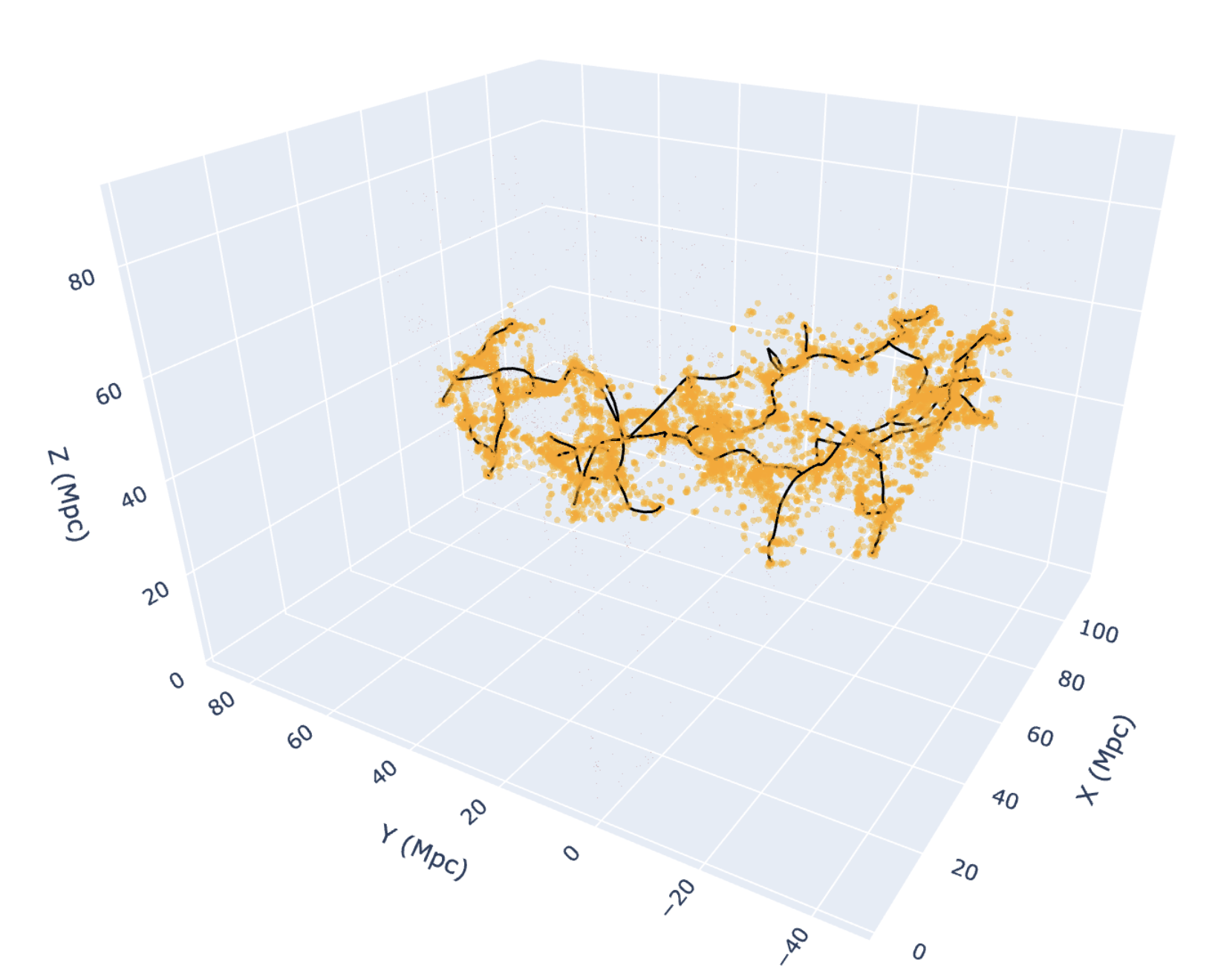}
\caption{Filament traversal density grid. Voxel intensity reflects the number of times it is crossed by a filament segment across all realisations.}
\label{fig:filament_density_grid}
\end{figure}

A three-dimensional filament density grid is constructed by counting the number of filament crossings in each voxel across all realisations. This grid, which serves as input for a final run of \texttt{DisPerSE} with the same parameters, is displayed in Fig.~\ref{fig:filament_density_grid}.

\begin{figure}
\centering
\includegraphics[width=0.4\textwidth]{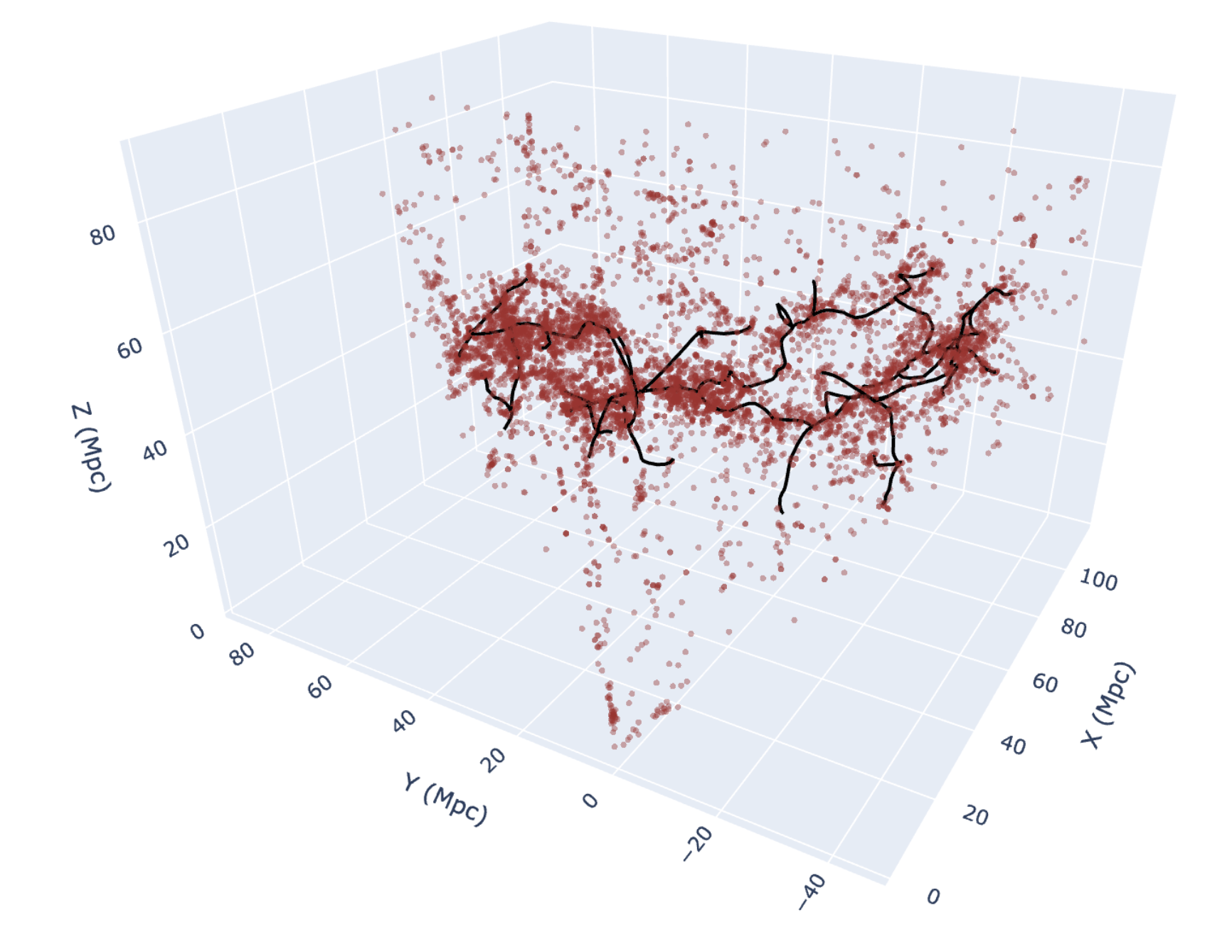}
\caption{Final skeleton extracted from the filament density grid: dark segments represent the structure and red dots correspond to galaxies.}
\label{fig:final_skeleton}
\end{figure}

The final skeleton is visually inspected, and segments located near the boundaries of the grid, typically isolated and likely affected by edge effects are removed. The cleaned result is shown in Fig.~\ref{fig:final_skeleton}.

\section{Description of identified clusters and groups}\label{app:clgr}

Tables~\ref{tab:clusters_perseus} and \ref{tab:clusters_sr} list the main galaxy clusters and groups identified within the CFHT-observed regions, region A et B respectively. For each system, the table reports its name, equatorial coordinates (RA, Dec) and redshift of the central galaxy corrected from FoG, as well as the characteristic radius $r_{200}$. Note that the typical relative uncertainty on $r_{200}$ is about 10\%, while redshift measurements carry uncertainties of the order of $5\times10^{-4}$.

\begin{table}
\centering
\begin{tabular}{lccccc}
\hline\hline
Name & RA($^\circ$) & Dec($^\circ$) & $z_{init}$ & $z_{corr}$ & $r_{200}$(Mpc) \\
\hline
NGC 996          & 39.666 & 41.648 & 0.0144 & 0.0180 & 0.616 \\
A 426            & 49.950 & 41.515 & 0.0172 & 0.0179 & 2.201 \\
UGC 1841         & 35.798 & 42.992 & 0.0209 & 0.0187 & 0.919 \\
AWM7             & 43.614 & 41.580 & 0.0172 & 0.0183 & 0.857 \\
HDCE 137         & 36.352 & 42.036 & 0.0178 & 0.0182 & 1.429 \\
HDCE 226         & 51.267 & 40.691 & 0.0123 & 0.0177 & 0.293 \\
\hline
\end{tabular}
\caption{Galaxy clusters and groups in region A.}
\label{tab:clusters_perseus}
\end{table}

\begin{table}
\centering
\begin{tabular}{lccccc}
\hline\hline
Name & RA($^\circ$) & Dec($^\circ$) & $z_{init}$ & $z_{corr}$ & $r_{200}$(Mpc) \\
\hline
WBL007           & 4.598  & 30.063 & 0.0227 &0.0226 & 0.763 \\
NGC 383          & 16.821 & 32.407 & 0.0170 & 0.0170 & 0.817 \\
NGC 410          & 17.743 & 33.152 & 0.0174 & 0.0172 & 0.295 \\
NGC 499          & 20.797 & 33.460 & 0.0147 & 0.0166 & 0.783 \\
NGC 507          & 20.916 & 33.256 & 0.0166 & 0.0168 & 1.023 \\
A 262            & 28.194 & 36.152 & 0.0162 & 0.0162 & 1.264 \\
NGC 777          & 30.062 & 31.429 & 0.0167 & 0.0161 & 0.736 \\
HDCE 1217        & 343.571 & 32.252 & 0.0218 & 0.0217 & 0.479 \\
HDCE 1243        & 352.146 & 32.416 & 0.0167 & 0.0171 & 0.648 \\
HDCE 23          & 9.622  & 29.511 & 0.0180 & 0.0187 & 0.312 \\
HDCE 43          & 15.309 & 30.131 & 0.0225 & 0.0233 & 0.344 \\
HDCE 103         & 29.226 & 33.044 & 0.0150 & 0.0151 & 0.290 \\
HDCE 105         & 29.537 & 33.203 & 0.0170 & 0.0153 & 0.291 \\
NGC 7831         & 1.831 & 32.609 & 0.0169 & 0.0166 & 0.289 \\
\hline
\end{tabular}
\caption{Galaxy clusters and groups in region B.}
\label{tab:clusters_sr}
\end{table}

\section{Examples of interactions and morphologies}\label{app:examples_interac_morpho}

\begin{figure}
\centering
\includegraphics[width=\columnwidth]{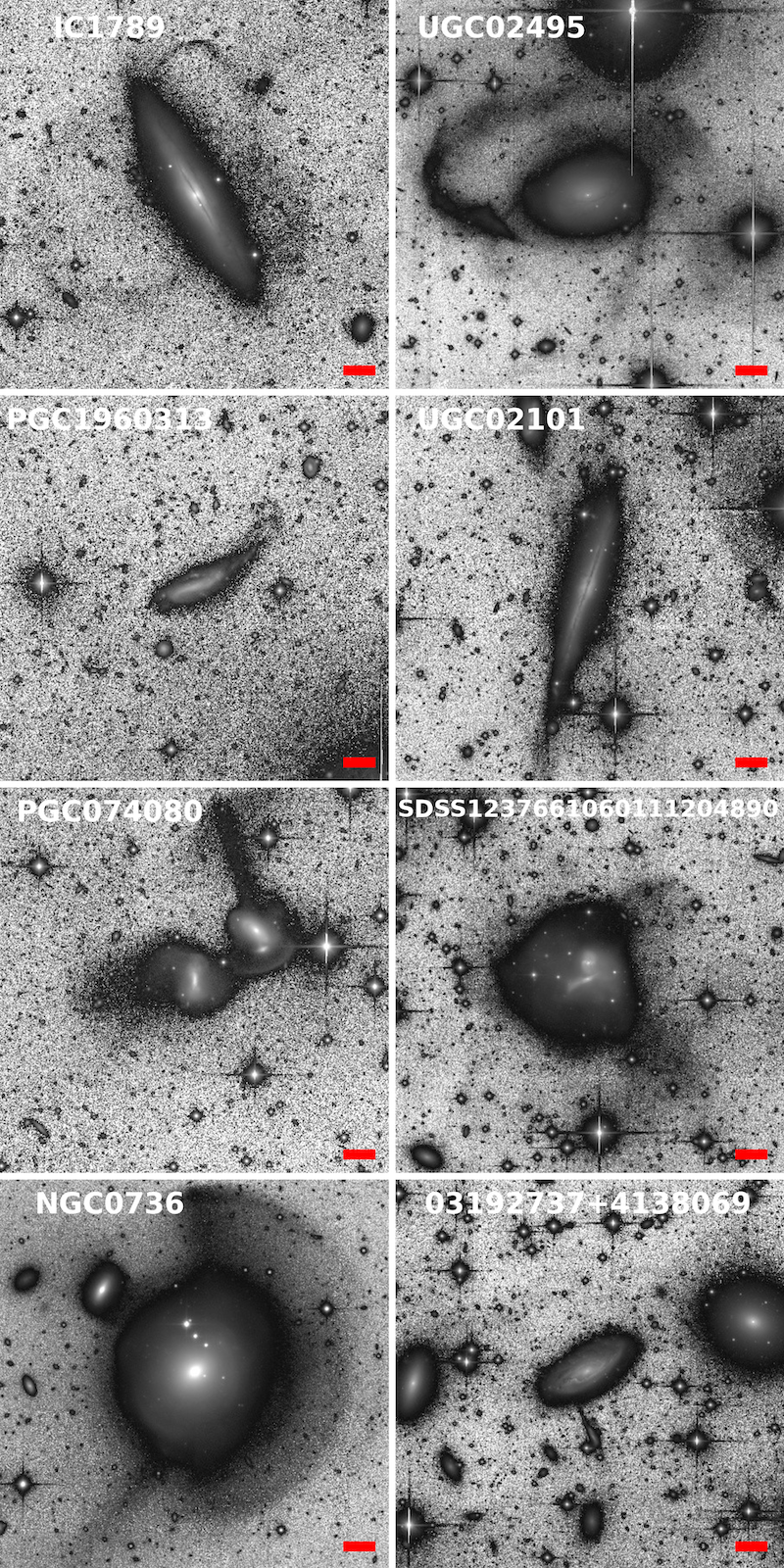}
\caption{Representative examples of interaction types classified by visual inspection and Sérsic profile shape. From top to bottom, the rows correspond to tidal tails, asymmetries, mergers, and shells or possible ram pressure stripping. The left column shows examples from the region A, and the right column from the region B. The red line in the bottom right corner of each panel represents a scale of 10 arcseconds.}
\label{interaction_examples}
\end{figure}

\begin{figure}
\centering
\includegraphics[width=\columnwidth]{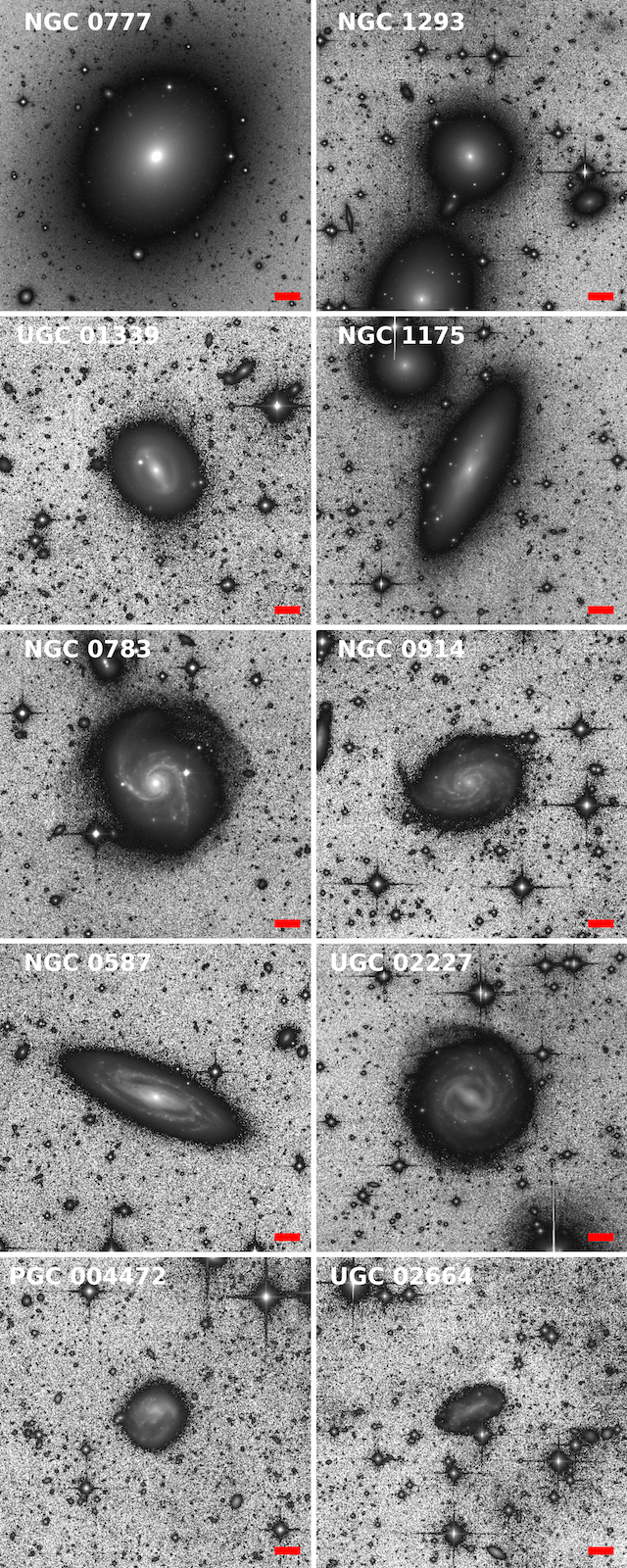}
\caption{Representative examples of morphological types classified by visual inspection and Sérsic profile shape. From top to bottom, the rows correspond to elliptical, lenticular (S0), spiral (S(B)a--S(B)d), and irregular (Irr) galaxies. The left (resp. right) column shows examples from region A (resp. region B).}
\label{morpho_examples}
\end{figure}

This appendix presents representative examples of the types of interactions and morphologies observed in our sample. Figures \ref{interaction_examples} and \ref{morpho_examples} illustrate, respectively, different galaxy interactions and morphological types, classified both by visual inspection and Sérsic profile shape.

\section{Catalogue and scaling relation}\label{app:sclingrelations}

\subsection{The PPSC catalogue}

This appendix presents the tables of galaxies parameters from PPSC catalogue, which list for each galaxy all measured parameters together with their corresponding uncertainties.  
Both the parameters and their associated errors are included within the same catalogue table for convenience. The complete table is available at the Strasbourg Astronomical Data Centre (CDS).

\begin{table*}

\begin{tabular}{lll}
\hline\hline
Parameter & Error parameter & Description \\
\hline
id                        &  & Unique galaxy ID \\
ra                        & ra\textsubscript{err} & Right Ascension [deg] \\
dec                       & dec\textsubscript{err} & Declination [deg] \\
z\textsubscript{init}     & z\textsubscript{init,err} & Initial redshift \\
z\textsubscript{corr}     & z\textsubscript{corr,err} & Corrected redshift after FoG reduction \\
mag                       & mag\textsubscript{err} & Apparent magnitude in CFHT-r band \\
mag\textsubscript{abs}    & mag\textsubscript{abs,err} & Absolute magnitude in CFHT-r band \\
R\textsubscript{e,pix}    & R\textsubscript{e,pix,err} & Effective radius [pixels] \\
R\textsubscript{e,arcsec} & R\textsubscript{e,arcsec,err} & Effective radius [arcsec] \\
R\textsubscript{e,kpc}    & R\textsubscript{e,kpc,err} & Effective radius [kpc] \\
R\textsubscript{25,pix}   & R\textsubscript{25,pix,err} & Radius at 25 mag/arcsec$^{2}$ [pixels] \\
R\textsubscript{25,arcsec}& R\textsubscript{25,arcsec,err} & Radius at 25 mag/arcsec$^{2}$ [arcsec] \\
R\textsubscript{25,kpc}   & R\textsubscript{25,kpc,err} & Radius at 25 mag/arcsec$^{2}$ [kpc] \\
Sersicn                       & Sersicn\textsubscript{err} & Sersic index \\
Axisratio                 & Axisratio\textsubscript{err} & Mean axis ratio \\
PA                        & PA\textsubscript{err} & Mean position angle [deg] \\
Axisratio\textsubscript{25} & Axisratio\textsubscript{25,err} & Axis ratio at 25 mag/arcsec$^{2}$ \\
PA\textsubscript{25}      & PA\textsubscript{25,err} & Position angle at 25 mag/arcsec$^{2}$ [deg] \\
sb\textsubscript{0}       & sb0\textsubscript{err} & Central surface brightness [mag/arcsec$^{2}$] \\
sbre                      & sbre\textsubscript{err} & Effective surface brightness [mag/arcsec$^{2}$] \\
sbre\textsubscript{avg}   & sbreavg\textsubscript{err} & Average effective surface brightness [mag/arcsec$^{2}$] \\
magaper1Re\textsubscript{CFHT-r} & magaper1Re\textsubscript{CFHT-r,err} & CFHT/MagaCam r-band magnitude within 1\,R\textsubscript{e} (CFHT-r) \\
magaper1Re\textsubscript{CFHT-r,abs} & magaper1Re\textsubscript{CFHT-r,abs,err} & Absolute CFHT/MagaCam r-band magnitude within 1\,R\textsubscript{e} \\
extcorr\textsubscript{CFHT-r} & extcorr\textsubscript{CFHT-r,err} & Extinction correction in CFHT/MagaCam r-band \\
Morphology                &  & Morphological type: Irregular, Spiral, Elliptical, Lenticular, \\
               &  & Merger, Asymmetry, Tidal tail, Shell \\
log\,M\textsubscript{*}/M$_\odot$ & log\,M\textsubscript{*}/M$_\odot$\textsubscript{err} & Log stellar mass [M$_\odot$] \\
objID\textsubscript{SDSS} &  & SDSS object ID \\
umag\textsubscript{SDSS}     & e\_umag\textsubscript{SDSS} & u-band magnitude from SDSS \\
gmag\textsubscript{SDSS}     & e\_gmag\textsubscript{SDSS} & g-band magnitude from SDSS \\
rmag\textsubscript{SDSS}     & e\_rmag\textsubscript{SDSS} & r-band magnitude from SDSS \\
imag\textsubscript{SDSS}     & e\_imag\textsubscript{SDSS} & i-band magnitude from SDSS \\
z\textsubscript{SDSS}     & e\_zmag\textsubscript{SDSS} & z-band magnitude from SDSS \\
z\textsubscript{sp,SDSS}  & e\_z\textsubscript{sp,SDSS} & Spectroscopic redshift from SDSS \\
z\textsubscript{ph,SDSS}  & e\_z\textsubscript{ph,SDSS} & Photometric redshift from SDSS \\
\hline
\end{tabular}
\caption{Summary of the parameters extracted for the PPSC galaxy catalogue, with their corresponding error parameters and descriptions.}
\label{tab:parameter_summary}
\end{table*}

\subsection{Scaling relation}

\begin{figure*}
\centering
\includegraphics[width=\textwidth]{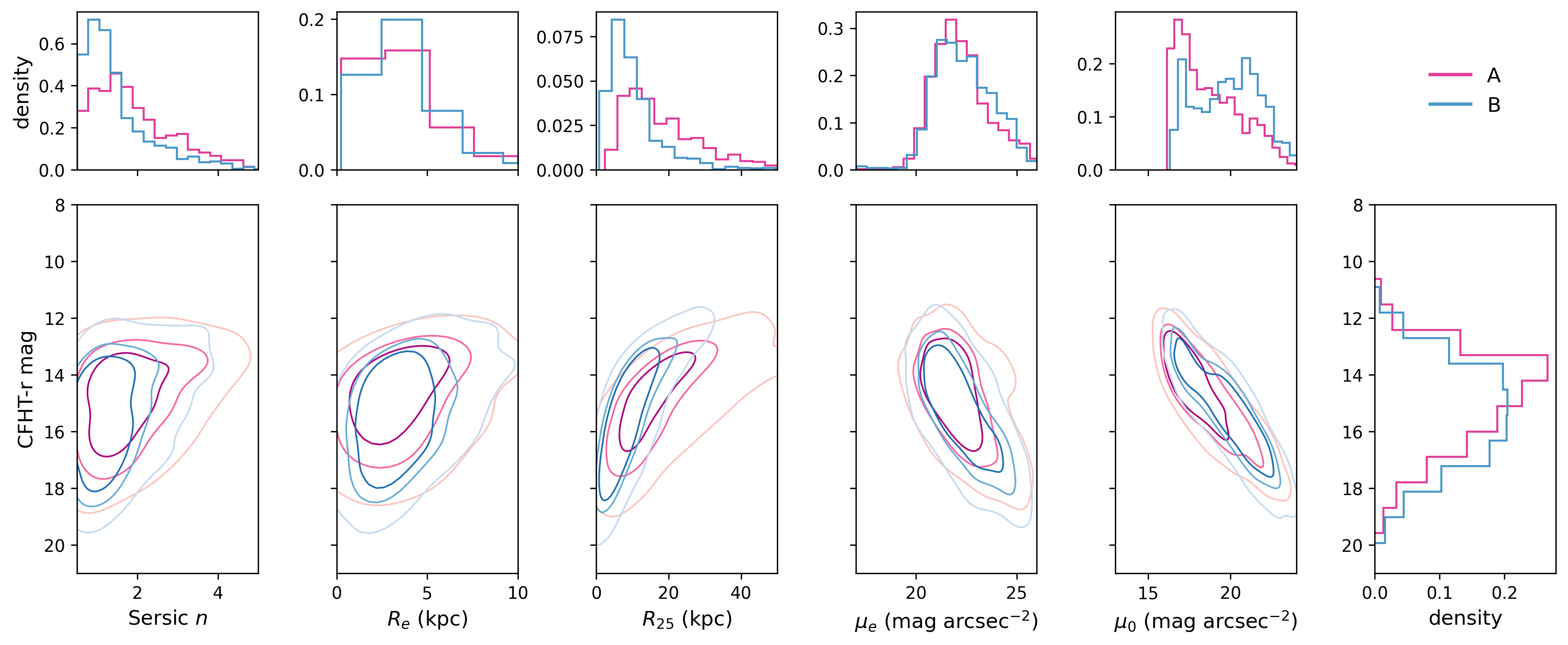}
\caption{Scaling relations of our PPSC sample: $r$-band magnitude versus Sérsic index $n$, effective radius $R_e$, effective surface brightness $\mu_e$, and central surface brightness $\mu_0$. Shaded contours indicate the 50\%, 1$\sigma$, and 2$\sigma$ levels of the total 2D Gaussian distributions, from darkest to lightest colours. Pink contours correspond to the region A, while blue contours represent the region B. One-dimensional histograms of key structural parameters are shown in the top row for the Sérsic index $n$, effective radius $R_e$, $\mu_e$, and $\mu_0$, and in the rightmost panel of the second row for the CFHT $r$-band magnitude.}
\label{scaling_relations}
\end{figure*}

To visualize the diversity of galaxy structural properties across environments, we present in Fig.~\ref{scaling_relations} the scaling relations between total $r$-band magnitude, Sérsic index $n$, effective surface brightness $\mu_e$, effective radius $R_e$, central surface brightness $\mu_0$, and stellar mass. These relations allow us to investigate how the global galaxy structure varies with environment. The distributions differ significantly: galaxies in the region B tend to have fainter magnitudes, smaller radii, and brighter central surface brightnesses compared to those in the region A.

\section{Stellar mass calibration and validation}\label{app:mass}

In this appendix, we provide a detailed description of the stellar mass calibration procedure and the validation of the resulting estimates.

\subsection{Crossmatch with SDSS photometry}

To obtain reliable colours for the majority of galaxies, we crossmatched our CFHT catalogue with the SDSS photometric database \citep[DR16;][]{SDSSDR16}. Approximately 50\% of our galaxies have a secure SDSS counterpart, providing $g$- and $r$-band magnitudes. These magnitudes were used to derive $g-r$ colours and to compute the mass-to-light ratio following \citet{Bell2003} with a Chabrier IMF \citep{Chabrier2003}.  

\begin{figure}
\centering
\includegraphics[width=0.5\textwidth]{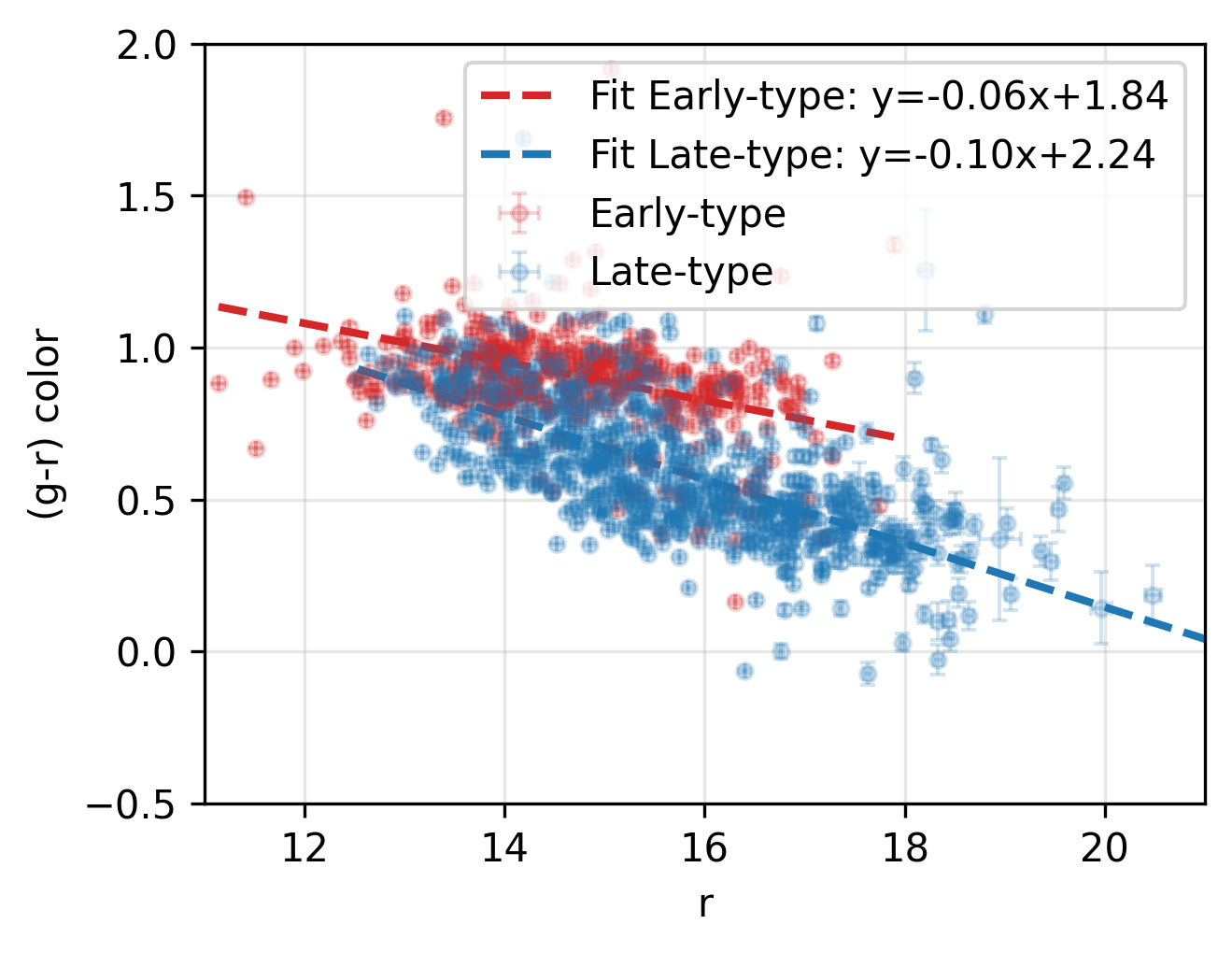}
\caption{$g-r$ colour as a function of $r$ magnitude for early-type (red dots) and late-type (blue dots) galaxies. Points are coloured by morphology, and the solid dashed lines indicate the linear relations used to estimate $g-r$ for galaxies without SDSS photometry.}
\label{fig:app_LTET}
\end{figure}

For galaxies lacking SDSS photometry, we first calibrated the CFHT $r$-band magnitudes onto the SDSS $r$ system using the relation derived from the SDSS–CFHT crossmatched sample. Their $g-r$ colours were then estimated using the linear relations between $g-r$ and $r$ magnitude obtained from SDSS–CFHT matched galaxies (Fig.\ref{fig:app_LTET}), separately for early-type and late-type systems and applied according to their morphological classification. These estimated colours were subsequently used to compute stellar masses in a consistent manner across the full sample. This procedure provides homogeneous stellar mass estimates, while introducing an uncertainty of roughly 0.1~dex.

\subsection{Validation of the mass calibration}

\begin{figure}
\centering
\includegraphics[width=0.5\textwidth]{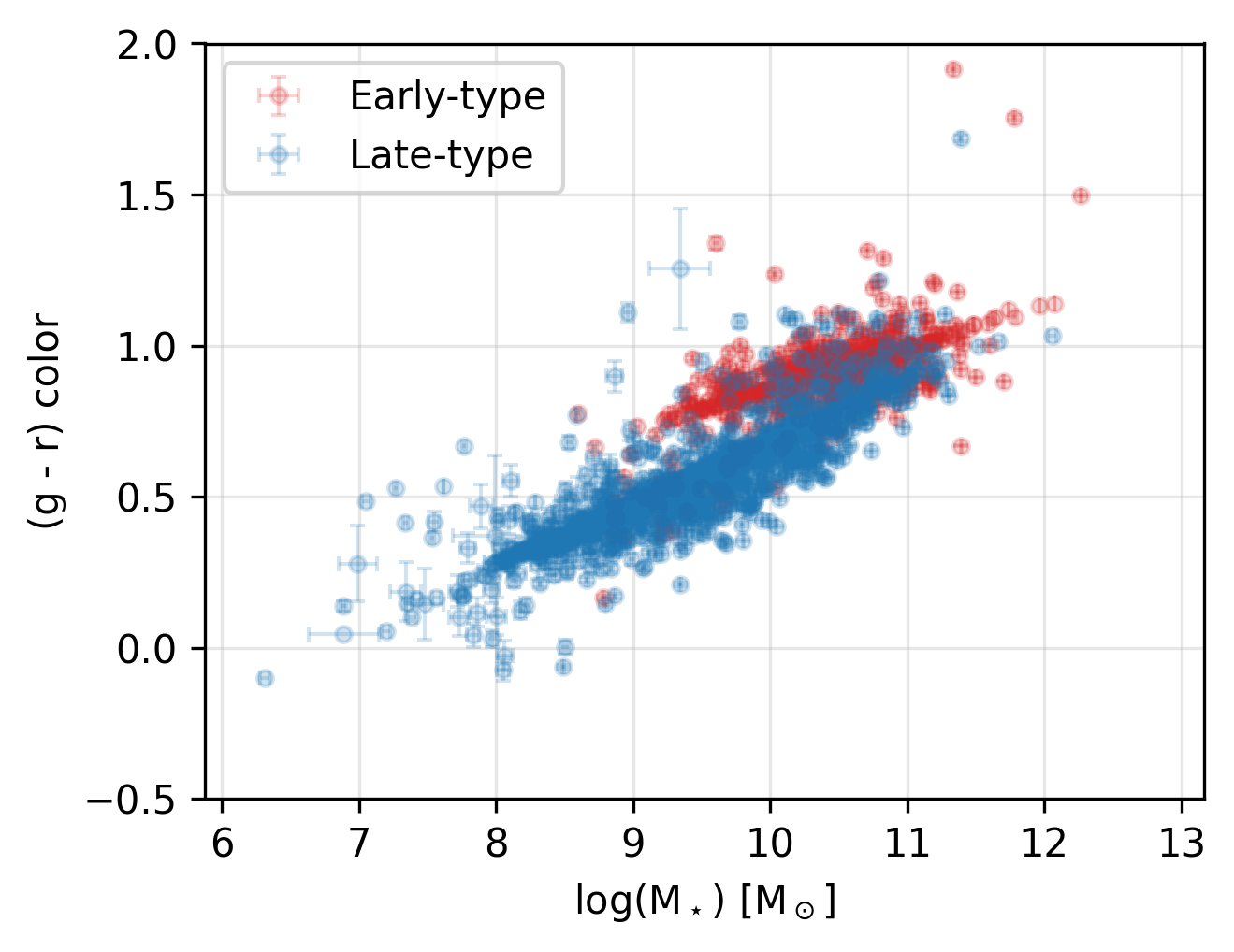}
\caption{$g-r$ colour versus stellar mass for different morphological types. Points are colour-coded by morphology: early-type in red and late-type in blue}
\label{fig:app_massA}
\end{figure}

\begin{figure}
\centering
\includegraphics[width=0.4\textwidth]{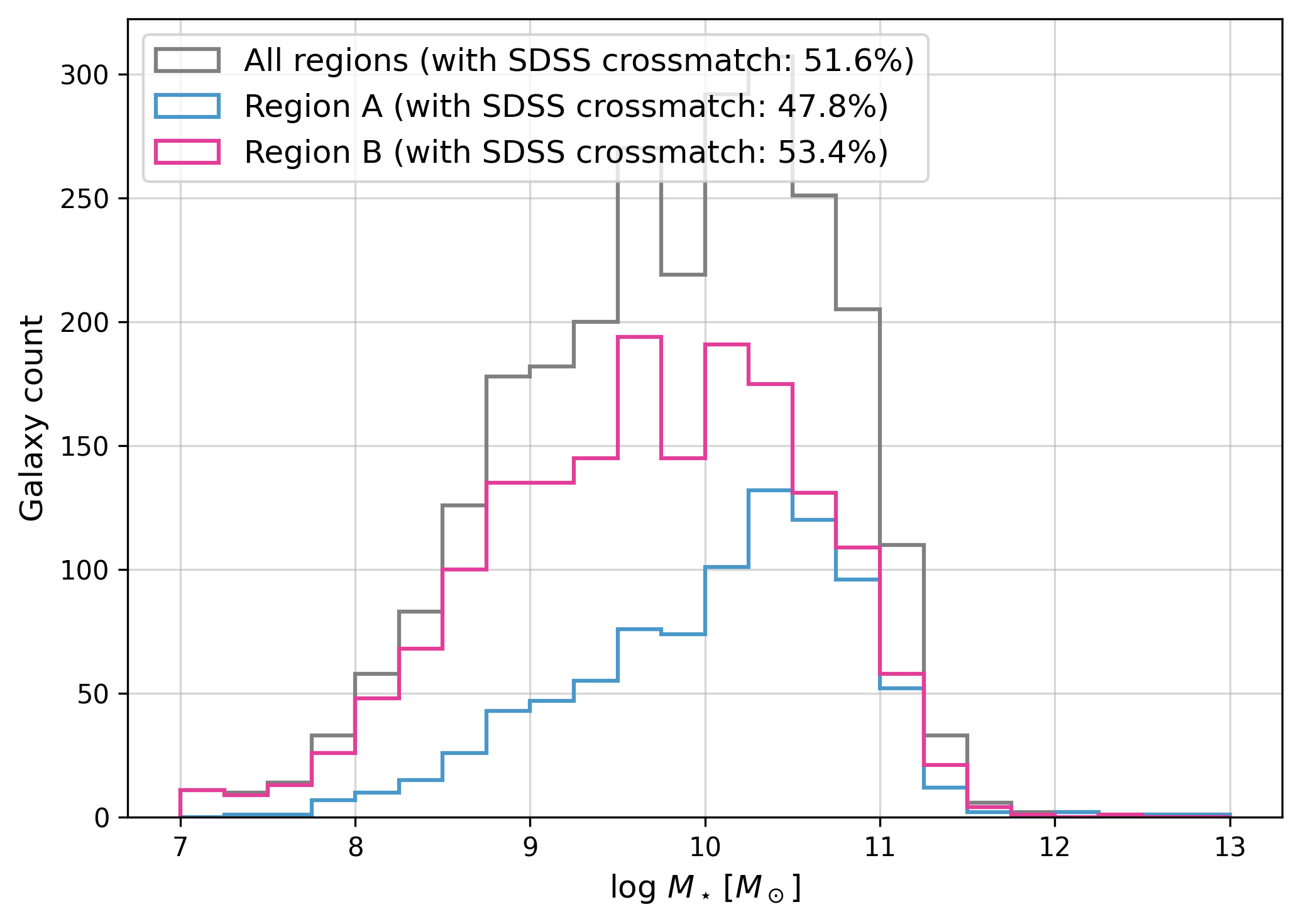}
\caption{Stellar mass distribution of the full galaxy sample and for each region}
\label{fig:app_massB}
\end{figure}

Several diagnostic plots were produced to validate the mass estimation procedure. For each morphological class, we examined the relation between $g-r$ colour and derived stellar mass Fig.~\ref{fig:app_massA}. The resulting distribution of stellar masses across the sample is also plotted in Fig.\ref{fig:app_massB}.


\section{Density profile of filaments}\label{app:filprofiles}

This appendix provides methodological details complementing Sect.~\ref{sec:filament_size}.

\subsection{Radial profiles of filaments}

The characteristic filament radius $r_{\mathrm{fil}}$ is derived from the radial galaxy density profiles following \citet{GalarragaEspinosa2020} and \citet{Wang2024}. 
We compute the normalized galaxy density $\rho/\rho_0$ in logarithmic annuli of perpendicular distance to the filament spine, excluding galaxies within $d_{\mathrm{cl}}$ of nearby clusters to minimize contamination. 
Profiles are modelled with a decaying half-Gaussian function; initial fits provide starting values that are explored with an MCMC sampler to quantify uncertainties. 
The logarithmic slope $\gamma = d\log_{10}\rho / d\log_{10} d_{\mathrm{fil}}$ is then derived from the best-fit model, and $r_{\mathrm{fil}}$ is defined as the position of the minimum of $\gamma$.  

\begin{figure}
\centering
\includegraphics[width=0.5\textwidth]{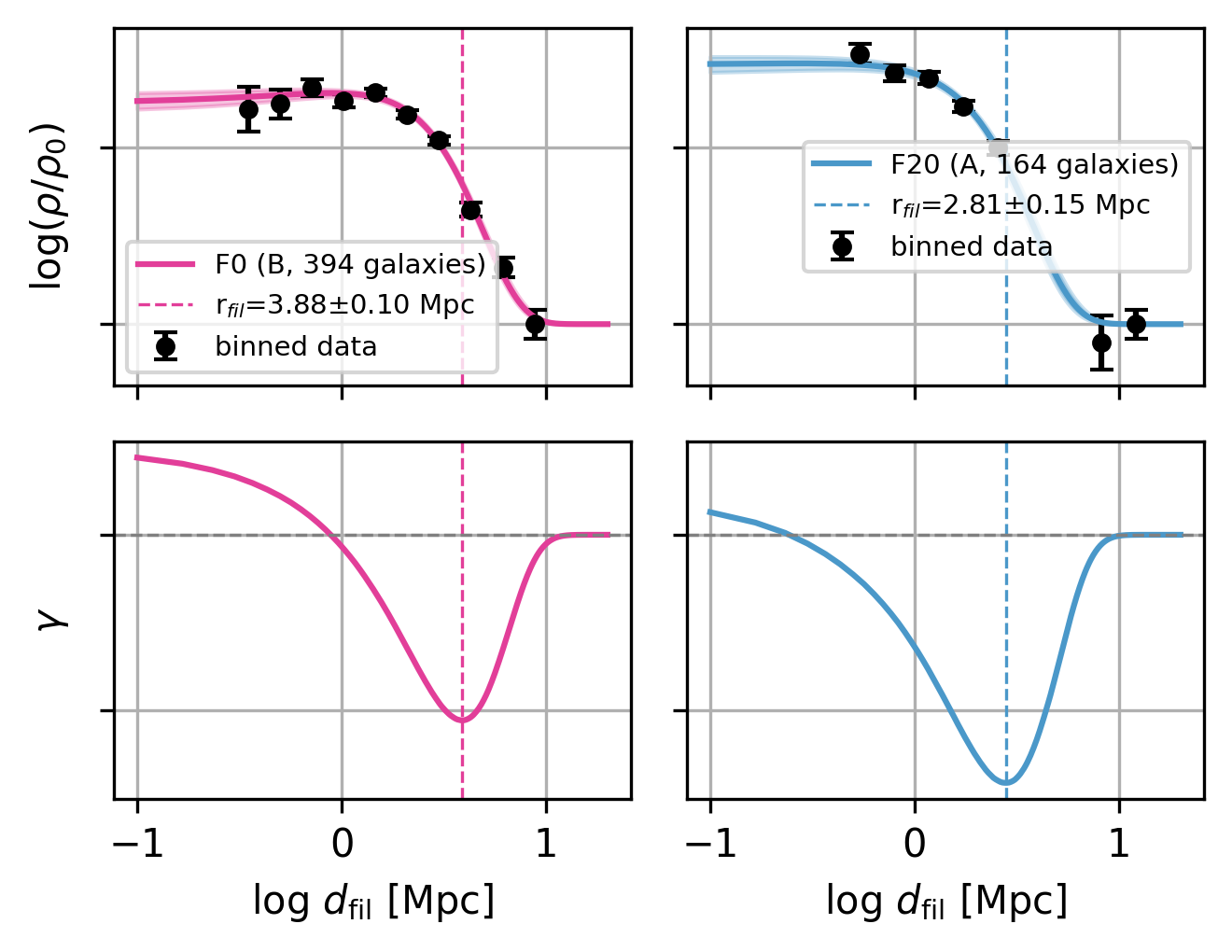}
\caption{
Examples of radial galaxy density profiles and their derivatives fitted with the half-Gaussian model. 
Top panels: binned galaxy densities with best-fit model and $1\sigma$ band. 
Bottom panels: slope $\gamma$, with the filament radius $r_{\mathrm{fil}}$ indicated at its minimum.
}
\label{fig:rfil_examples}
\end{figure}

\subsection{Galaxy counts and mass assignment}

Galaxies with $M_\star > 10^{8.5}\,M_\odot$ are assigned to clusters, filaments, or outskirts based on their distance to the nearest cluster center and filament spine. 
The adopted boundary radii are $r_\mathrm{200}$ Mpc and $r_\mathrm{fil}$ Mpc.  

Stellar masses are corrected for incompleteness and converted to halo masses using a fixed stellar-to-halo conversion factor $f_\star = 0.015$ \citep{Moster2013, Behroozi2013}.  

Table~\ref{tab:galaxy_counts} summarizes the galaxy counts per environment, while Table~\ref{tab:stellar_mass} lists the stellar and halo mass budgets.  

\begin{table}[h!]
\centering
\begin{tabular}{lccc}
\hline\hline
Region & Clusters & Groups & Filaments\\
\hline
A & 98 & 15 & 555 \\
B & 46 & 33 & 949 \\
\hline
\end{tabular}
\caption{Galaxy counts in different cosmic web environments. Membership is defined using the adopted $r_{\mathrm{200}}$ and $r_{\mathrm{fil}}$ thresholds.}
\label{tab:galaxy_counts}
\end{table}

\begin{table}[h!]
\centering
\begin{tabular}{lccc}
\hline\hline
Region & Sub-region & Stellar mass & Halo mass\\
 &  & ($\times 10^{13}M_\odot$) & ($\times 10^{14}M_\odot$) \\
\hline
A& Clusters/Groups & $(1.02\pm0.41)$  & $(6.83\pm0.94)$ \\
& Filaments & $(2.79\pm0.31)$ & $(18.5\pm0.20)$ \\
\hline
B & Clusters/Groups & $(0.60\pm0.09)$ & $(3.99\pm0.10)$ \\
& Filaments & $(2.71\pm0.29)$ & $(18.1\pm0.19)$ \\
\hline
\end{tabular}
\caption{Stellar and halo mass estimates in the PPSC subregions, corrected for incompleteness and scaled by $f_\star = 0.015$.}
\label{tab:stellar_mass}
\end{table}

\section{Fraction details}\label{app:fraction_details}

This appendix provides complementary analyses to the main results, focusing on detailed variations of morphological and interaction fractions as a function of the normalised distance to the main environmental structures: the filament spine ($d_{\mathrm{fil}}/r_{\mathrm{fil}}$) and the centres of groups and clusters ($d_{\mathrm{gr,cl}}/r_\mathrm{200}$). Figures~\ref{fig:morpho_fraction_detailed}, \ref{fig:interactionsdetail}, and \ref{fig:combined_interactions} illustrate these trends for the different environments (Filaments, Groups, and Clusters).

\subsection{Morphology as a function of distance to filament, group, and cluster}

\begin{figure*}[htbp]
\centering
\includegraphics[width=0.7\textwidth]{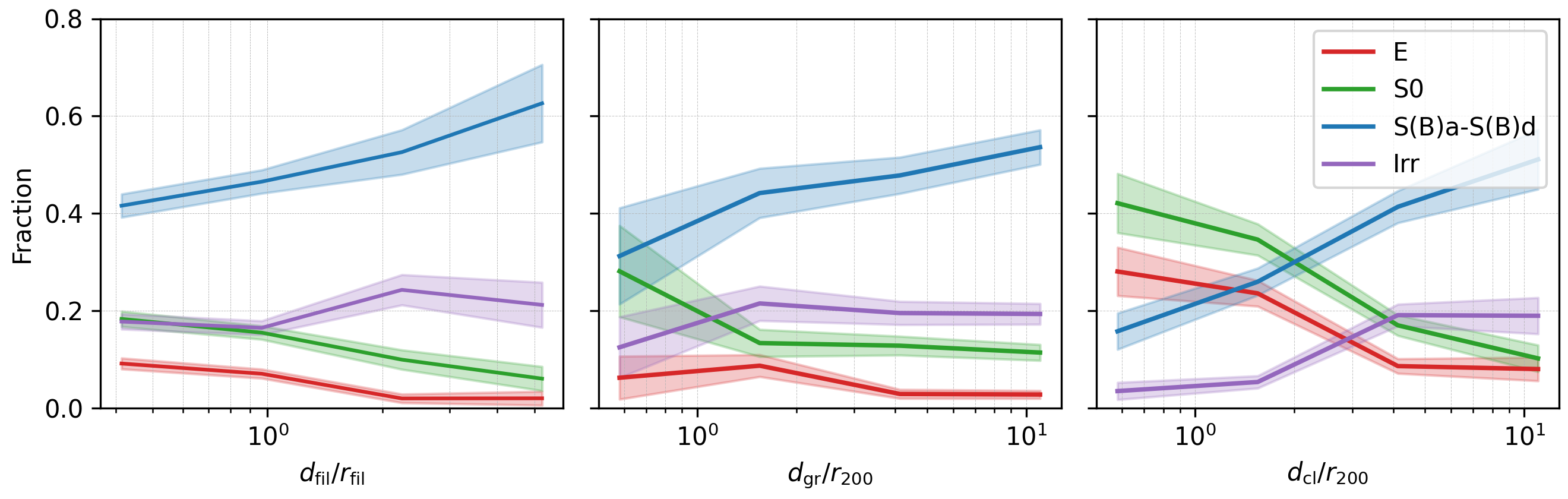}
\caption{
Morphological fractions versus normalised distance to the filament ($d_\mathrm{fil}/r_\mathrm{fil}$, left), group centre ($d_\mathrm{gr}/r_\mathrm{200}$, middle), and cluster centre ($d_\mathrm{cl}/r_\mathrm{200}$, right). Colours indicate morphological types: E (red), S0 (green), S(B)a--S(B)d (blue), and Irr (purple). Shaded areas represent binomial uncertainties.
}
\label{fig:morpho_fraction_detailed}
\end{figure*}

Figure~\ref{fig:morpho_fraction_detailed} presents the fractions of galaxies by morphological type as a function of the normalised distance to the filament spine (left), to the nearest group (middle), and to the nearest cluster (right).

In filaments, S(B)a--S(B)d galaxies dominate at all radii, with a fraction rising from about $\sim0.4$ in the filament spine to higher values toward outskirts. S0 galaxies contribute $\sim0.2$, while ellipticals remain around $\sim0.1$ near the filament core. Irregular galaxies show a significant increase beyond the filament radius, suggesting that late-type and disturbed systems are more common in the outer, lower-density parts of filaments.

In groups, spirals still dominate but with a lower central fraction ($\sim0.3$) near the filament spine, accompanied by a comparable contribution of S0s. This pattern indicates a population of spiral galaxies undergoing morphological transformation and progressive quenching within the group potential. Beyond $r_\mathrm{200}$, irregular galaxies become more frequent than early types, consistent with milder environmental effects in the outskirts.

In clusters, the inversion of population is more pronounced. The spiral fraction drops to $\sim0.18$ at the cluster centre, while S0s dominate ($\sim0.4$) and ellipticals reach about $\sim0.4$, typical of dynamically evolved cores. Irregulars are nearly absent in the densest regions. Moving outward, spirals successively overtake ellipticals beyond $r_\mathrm{200}$ and S0s around $2\,r_\mathrm{200}$, while irregulars only exceed the early-type fractions at larger radii ($\sim4\,r_\mathrm{200}$). These trends trace the gradual transition from quiescent to star-forming populations with increasing clustercentric distance.

\subsection{Local interactions}

\begin{figure*}[htbp]
\centering
\includegraphics[width=0.8\textwidth]{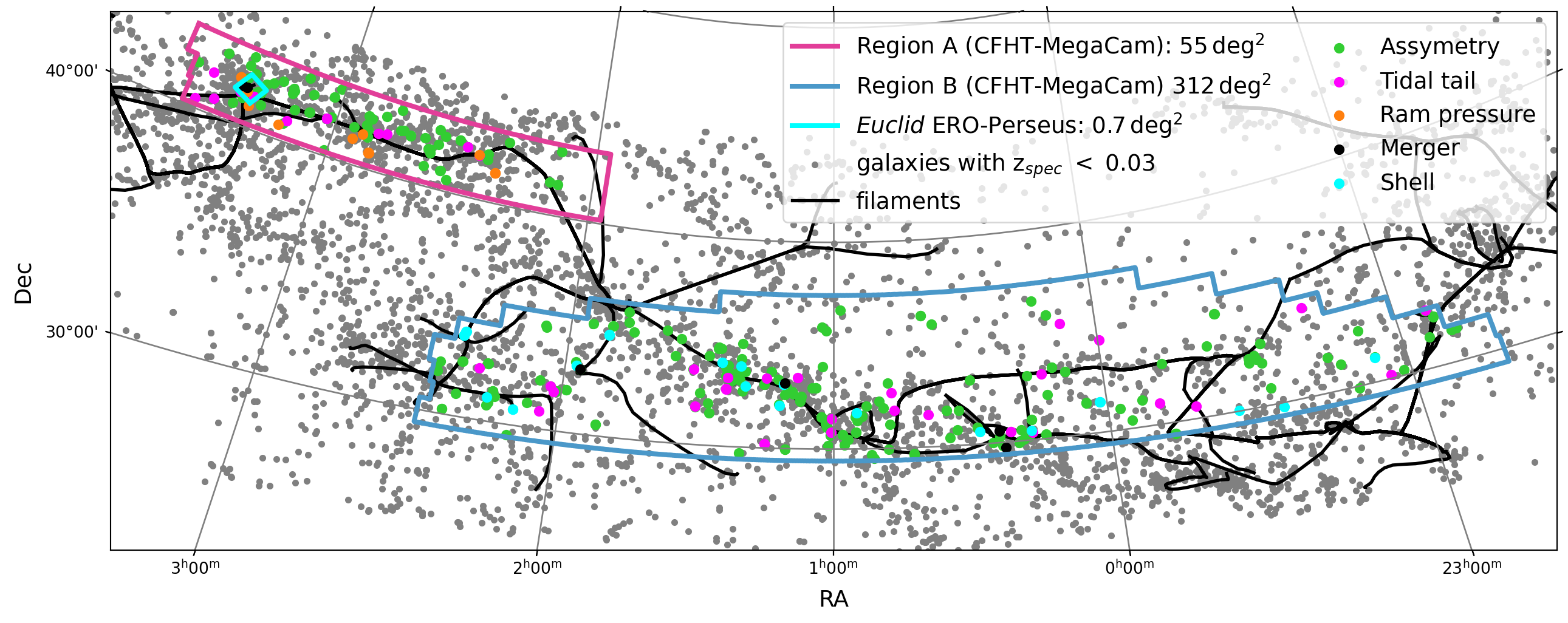}
\caption{
Projected distribution of galaxies by interaction type. Grey points show the global catalogue, coloured points indicate identified interaction features, and dark lines trace the projected 3D cosmic structure.
}
\label{fig:combined_interactions}
\end{figure*}

Figure~\ref{fig:combined_interactions} shows the projected spatial distribution of galaxies coloured by interaction type. The map highlights the concentration of disturbed systems along filaments and in the outskirts of massive haloes, tracing active sites of galaxy transformation.

\begin{figure*}[htbp]
\centering
\includegraphics[width=0.7\textwidth]{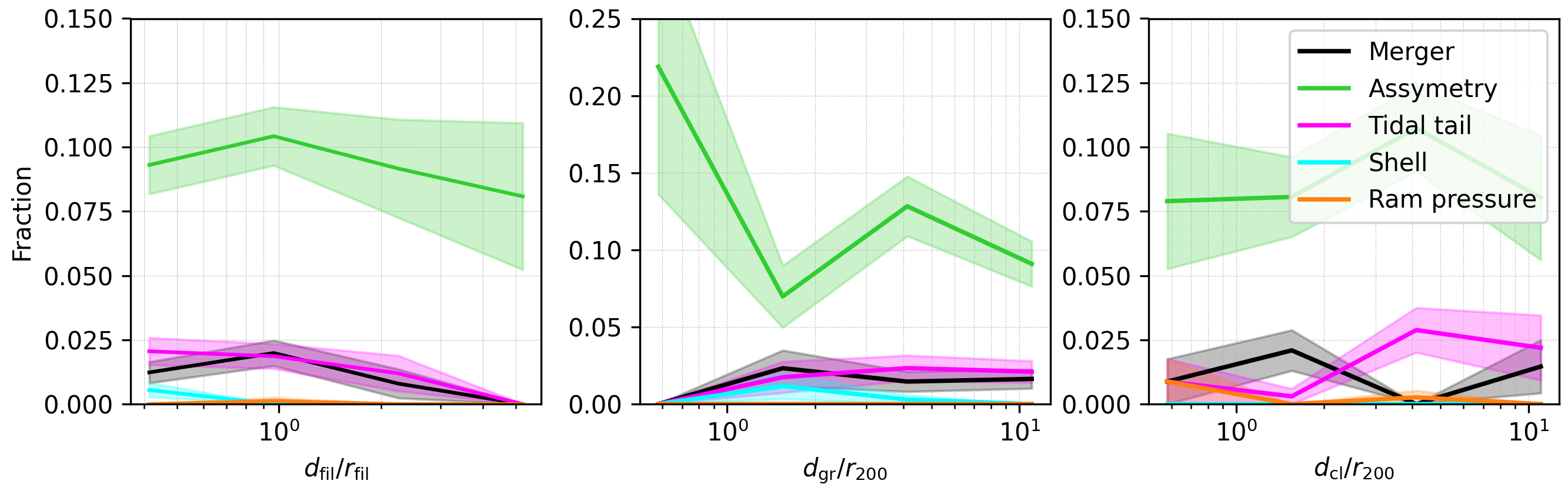}
\caption{
Interaction fractions as a function of characteristic distance: mergers (black), ram-pressure stripping effects (orange), tidal tails (magenta), shells (cyan), and halo asymmetries (green). Left: $d_\mathrm{fil}/r_\mathrm{fil}$; middle: $d_\mathrm{gr}/r_\mathrm{200}$; right: $d_\mathrm{cl}/r_\mathrm{200}$.
}
\label{fig:interactionsdetail}
\end{figure*}

Figure~\ref{fig:interactionsdetail} quantifies the interaction fractions by type as a function of distance to the filament, group, and cluster centres.

Across all environments, halo asymmetries represent the dominant category, as they encompass a broad range of low-surface-brightness disturbances not classified in other categories.  

In filaments, asymmetries and mergers are relatively common within the filament radius and show a mild increase toward the filament core, then decline outward. Tidal tails and shells decrease steadily with increasing distance, consistent with weaker tidal fields in the outskirts.

In groups, asymmetries remain significant within$r_\mathrm{200}$ (fraction $>0.2$) and drop sharply beyond this limit. Other interaction features, such as tidal tails and shells, peak just outside $r_\mathrm{200}$ before slowly declining, suggesting recent encounters and mergers in the group periphery.

In clusters, the overall interaction rate is low. Central regions show a small excess of ram-pressure and asymmetric features, while the outskirts (beyond $r_\mathrm{200}$) exhibit a modest rise in all interaction types, including some mergers and tidal tails, before declining again at larger distances. This behaviour reflects the complex interplay between high relative velocities, dense intracluster gas, and recent infall from surrounding filaments.

\end{appendix}

\end{document}